\documentclass[a4paper,10pt]{article}
\usepackage{amsmath}
\usepackage{a4wide}
\usepackage{bm}
\usepackage{graphicx}
\usepackage{subfig}
\newcommand{\SU}[1]{\mathrm{SU}(#1)}
\newcommand{\U}[1]{\mathrm{U}(#1)}
\newcommand{\Tr}[2]{\mathop{\mathrm{Tr}}_{#1}\left[#2\right]}

\newcommand{\nbrack}[1]{\left(#1\right)}
\newcommand{\sbrack}[1]{\left[#1\right]}
\newcommand{\tbrack}[1]{\left<#1\right>}

\newcommand{\sep}[1]{\hspace{5mm}\mbox{#1}\hspace{5mm}}
\newcommand{\brm}[1]{\bm{\mathrm{#1}}}
\newcommand{\cbrm}[1]{\overline{\bm{\mathrm{#1}}}}
\newcommand{\widerow}{\rule{0pt}{2.5ex}\rule[-1.5ex]{0pt}{0pt}}
\newcommand{\be}{\begin{equation}}
\newcommand{\ee}{\end{equation}}
\newcommand{\ba}{\begin{eqnarray}}
\newcommand{\ea}{\end{eqnarray}}
\numberwithin{equation}{section}
\numberwithin{figure}{section}
\numberwithin{table}{section}
\begin{document}

\date{\mbox{ }}
\title{{\normalsize  IPPP/09/16 DCPT/09/32\hfill\mbox{}\hfill\mbox{}}\\
\vspace{2.5 cm}
\Large{\textbf{Electric/Magnetic Duality with Gauge Singlets}}}
\vspace{2.5 cm}
\author{Steven Abel, James Barnard\\[3ex]
\small{\em Institute for Particle Physics Phenomenology, Durham University, Durham DH1 3LE, UK}\\[1.5ex] 
\small{\em }s.a.abel or james.barnard@durham.ac.uk\\[1.5ex] }
\date{}
\maketitle

\vspace{2ex}
\begin{abstract}
\noindent
We demonstrate how gauge singlets can be used to find new examples
of Kutasov duality (i.e. where the matching of the dual theories relies on 
a non-zero superpotential) in ${\cal N}=1$ $\SU{N}$ SQCD with $F_Q$ flavours of quark
and  multiple generations of adjoints, or antisymmetrics, or 
symmetrics. The role of the singlets is to simplify greatly 
the truncation of the chiral ring whilst maintaining an $R$-symmetry, and at the same time 
allowing an unambiguous identification of the elementary mesons of the magnetic theory. The dual 
theories satisfy all the usual tests, including 
the highly non-trivial 't~Hooft anomaly matching conditions. 
\end{abstract}

\newpage

\section{Introduction and background}

An interesting possibility in the context of electric-magnetic duality \cite{Intriligator:1995au} is that the supersymmetric Standard Model is the magnetic dual of some unknown electric theory \cite{Strassler:1996ua,Klebanov:2000hb,AK:Dualification}. This would have a bearing on many important questions. For example it would accommodate Landau poles in models of direct gauge mediation of supersymmetry (SUSY) breaking. The dynamical scales could also explain hierarchies in the SUSY breaking sector and even in the Yukawa couplings. More recently \cite{AK:Dualification} it was argued that this possibility may, rather counterintuitively, be compatible with unification. That argument was based on known examples of Kutasov duality, i.e. duality that relies on a non-zero superpotential \cite{K:Adj}. The vacuum structure of these models was analysed in the simplest SQCD plus a single adjoint case by Kutasov, Schwimmer and Seiberg (KSS); it can be that of a broken GUT theory with electric and magnetic dual descriptions \cite{KS:DKSS,KSS:DKSS}. Further related studies of these theories were made in Refs.\cite{Poppitz:1996wp,Poppitz:1996vh,Klein:1998uc,Klein:2003wa}.

\begin{figure}[!h]
\begin{center}
\includegraphics[width=75mm]{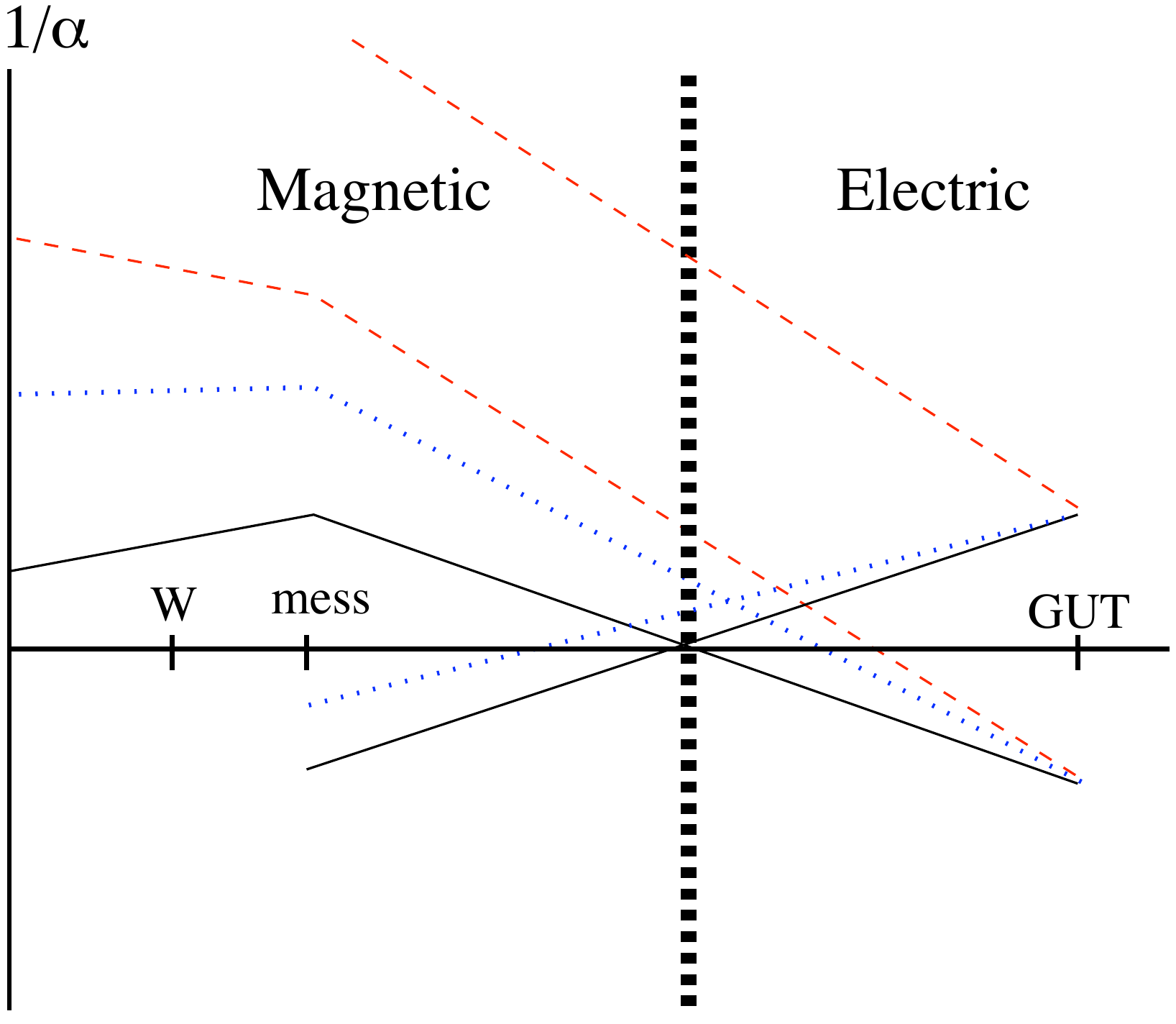}
\caption{\em The dual-unification scenario of Ref.\cite{AK:Dualification}: the supersymmetric Standard Model appears to run to unphysical gauge unification when there are many messengers in complete $SU(5)$ multiplets. This is mapped to a real unification occurring in an electric dual description that is  valid above the Landau pole scale. \label{fig:dualification}}
\end{center}
\end{figure}

The observation made in Ref.\cite{AK:Dualification} was that (as shown in Figure \ref{fig:dualification}) unification in such theories is preserved in the mapping from electric to magnetic descriptions, with the gauge couplings of the magnetic theory appearing to unify at the same energy scale but at unphysical (imaginary) values of the gauge couplings.  This unphysical gauge unification also happens in a supersymmetric Standard Model with a large number of messengers in complete $\SU{5}$ representations, and so in Ref.\cite{AK:Dualification} it was suggested that the latter would be a strong hint of a magnetic dual GUT theory.  Furthermore it was argued that this would explain why the supersymmetric Standard Model appears to unify, but the proton does not decay. The proton decay operator (being generated at the GUT scale) is a baryon operator that is generated in the electric theory. However the decay takes place at low energies and is therefore computed in the magnetic description with the relevant operator being mapped to the corresponding magnetic baryon: consequently the decay rate is suppressed by the many powers of $\Lambda/M_{GUT}$ associated with the mapping of electric to magnetic baryons.

Unfortunately the arguments of Ref.\cite{AK:Dualification} are limited by the fact that no convincing electric dual description of the supersymmetric Standard Model or indeed the Georgi-Glashow $\SU{5}$ model is known. In particular, in for example an $\SU{5}$ framework, one would like to be able to build chiral models, models with three generations, with adjoint GUT Higgs fields and antisymmetrics and so on. Although some of these ingredients (for example chiral models) are present in the models in the literature (related examples with $\SU{N}$ gauge groups can be found in Refs.\cite{KSS:DKSS,B:2Adj,BS:Theatre,ILS:NewDualities}), the particle content is very constrained and is far from that of the Georgi-Glashow model. Motivated by this fact, we will in this paper demonstrate how one can extend the Kutasov class of electric-magnetic dual theories using gauge and flavour singlets. 

In very general terms our approach can be described as follows. First recall the three basic tests of electric-magnetic duality \cite{S:Duality}: the two dual theories should
\begin{itemize}
\item have the same moduli space of vacua, and in particular a one-to-one 
map between the degrees of freedom represented by meson and baryon operators,
\item they should share the same global symmetries and have the same global anomalies (i.e. the 't~Hooft anomaly matching conditions),
\item their moduli spaces should still match under all deformations of the theories by chiral operators. 
\end{itemize}
The KSS-like theories can be shown to pass all these tests, thanks to the IR dynamics being simplified by the addition to the superpotential of various operators. In its  
original formulation, the KSS theory consisted of SQCD plus a single adjoint, $X$, and the 
superpotential operator
\be 
W\supset X^{k+1}
\ee
for some integer $k$. This term alters the infra-red dynamics (it is a dangerously irrelevant operator, as one says) for appropriate choices of $\SU{N}$ and $F_Q$ flavours, and truncates the chiral ring, allowing one to match the degrees of freedom in the moduli space relatively easily. Further examples of this kind of duality were discovered in Refs.\cite{BS:Theatre,ILS:NewDualities} involving more adjoint fields, and/or symmetric and antisymmetric representations.  An obstacle restricts the whole class of models however: as more fields are added, the chiral ring becomes more difficult to truncate and additional operators have to be added to the superpotential to do this; on the other hand each new operator projects out some global symmetry and weakens the 't~Hooft anomaly matching test. In particular once the $R$-symmetry is explicitly broken 
the 't~Hooft anomaly matching test becomes virtually meaningless, so this symmetry at least one needs to preserve. Our approach in order to overcome this problem is to use an additional singlet to enable us to add the extra operators needed to truncate the chiral ring while preserving a meaningful $R$-symmetry. The role of this singlet is to break the $R$-symmetry spontaneously by acquiring a VEV, thereby generating the $R$-symmetry breaking couplings necessary to truncate the chiral ring. 

At first sight one might suppose that there is little to be gained from doing this. Indeed the $R$-charged couplings (and hence the singlet) 
will generally appear in the elementary mesons that are a feature of the 
spectrum of the magnetic dual theory, so it might seem that the $R$-charges in the 
magnetic theory are going to be arbitrary, and the correct anomaly matching condition impossible to identify. This is not so however: the matching of the composite mesons of the electric description to the elementary mesons of the magnetic one entails an assignment of singlet contribution that is unambiguous. Moreover a nontrivial test is that the equivalent assignment in the baryons makes their  global charges match automatically. With the $R$-charges of the elementary mesons of the magnetic dual theory duly fixed, anomaly matching can then proceed as normal, and it provides the usual powerful test of the duality.

In the following section we begin by showing how this can be done in the original KSS (SQCD plus single-adjoint) model of Refs.\cite{KS:DKSS,KSS:DKSS}. An important feature of this model is as we have stated the possibility of adding deformations that would break the GUT symmetry. Although these operators are not crucial for truncating the chiral ring (being relevant deformations rather than drastically changing the 
IR behaviour as the original $X^{k+1}$ operator did), it is still interesting to see how they can be made compatible with a preserved $R$-symmetry by the addition of a gauge singlet. The procedure required to identify the elementary mesons of the magnetic description in this simple case will serve as a useful guide for later models. We find that anomaly matching in the deformed theory works in two cases; the first corresponds to the original deformed KSS model with the singlet VEV merely parameterizing the size of the deformation. The second model is different from the original model, and as we shall see is (unlike the first) not associated with the flow to a fixed point. 

In sections 3, 5 and 6 we consider models with two or more adjoints and models with extra antisymmetric and symmetric representations respectively. In each case we can add to the superpotential enough operators to truncate the chiral ring while still maintaining an $R$-symmetry and therefore highly non-trivial 't~Hooft anomaly matching conditions. 

Section 4 deals in more detail with the RG flow. Concentrating on the adjoint models, we construct 
a complete picture of the flow in models with one adjoint plus a singlet using 
$a$-maximization~\cite{IW:aMax,Barnes:2004jj}.  The additional singlet alters the RG flow, and gives various fixed points; the flow is to the KSS model with the singlet appearing in a deformation term, however the theory can flow between fixed points corresponding to different deformations.  
We use this picture to deduce the RG flow behaviour for the more general models with additional adjoints. For the models with an odd number of adjoint fields, we argue that the endpoint of the flow is also the deformed KSS model. Models with an even number of adjoints flow to normal SQCD. In both cases the duality at the endpoints of the flow is well understood. 

We close the Introduction by making a parenthetical remark about the phenomenology. Recall that our ultimate goal is to be able to describe the supersymmetric Standard Model (or a Georgi-Glashow like GUT theory) as a magnetic dual. These theories generally do not allow anomaly-free global $R$-symmetries because of the complexity of their Yukawa couplings and Higgs mass terms. It is therefore likely that 
any $R$-symmetry would have to be spontaneously broken anyway in order to have reasonable phenomenology, and using gauge singlets to do this would be the simplest way. We consider this to be an additional motivation for the study in this paper.

\section{$R$-symmetry in the deformed KSS model\label{sec:DKSS}}

Let us first revisit the model of KSS in order to see how singlets can be introduced. 
As we have said, one of the appealing features of those models was the ability to match the electric and magnetic theories under deformations of the electric superpotentials
\begin{equation}\label{eq:deform}
W_{\mathrm{el}}=\sum_{i=0}^{k-1}\frac{t_i}{k+1-i}\Tr{}{X^{k+1-i}}+\lambda\Tr{}{X}
\end{equation}
where $X$ is an adjoint field of the $\SU{N}$ gauge group and we have chosen our basis for $X$ such that there is no $X^k$ term as explained in Ref.\cite{KSS:DKSS}. The undeformed theory has only the $t_0$ term.  The $t_{i>0}$ deformations spontaneously break the gauge symmetry and lead to a rich vacuum structure; it is then possible to show that there exists a similar deformation in the magnetic theory that produces the same vacuum. There is an aspect of this procedure that we will address in this section as a warm-up exercise which is this.  By adding the extra terms in Eq.\eqref{eq:deform}, one breaks the $R$-symmetries of the model.  In principle therefore the 't~Hooft anomaly matching conditions apply only to the undeformed theory.  However it is often useful to think of couplings such as $t_{i>0}$ as background $R$-charged fields that acquire VEVs.  Thus a way to match the anomalies directly in the deformed theory would be to consider these fields as singlets in the spectrum and to do the anomaly matching on the 
complete theory. The singlet VEVs can then be fixed at the end to generate the required $R$-breaking terms in Eq.\eqref{eq:deform} spontaneously. This is what we will investigate here.

We commence by summarizing the models of KSS \cite{KS:DKSS,KSS:DKSS} in more detail.  They are based on an $\SU{N}$ gauge group with $F_Q$ flavours of quarks and anti-quarks, and an adjoint field of the $\SU{N}$ denoted by $X$. The symmetry content is
\begin{equation}
\SU{F_Q}_L\times\SU{F_Q}_R\times\U{1}_B\times\U{1}_{R_{1}}\times\U{1}_{R_{2}}\, .
\end{equation}
When there is only the $t_0$ term in the superpotential, the global symmetry is partially broken but retains a $\U{1}_R$ symmetry.  The matter content is then summarised by Table \ref{tab:KSSem}.
\begin{table}[!h]
\begin{equation}\nonumber
\begin{array}{|c|c|c|c|c|c|}\hline
\widerow & \SU{N} & \SU{F_Q}_L & \SU{F_Q}_R & \U{1}_B & \U{1}_R \\\hline
\widerow Q & \brm{N} & \brm{F_Q} & \brm{1} & \frac{1}{N} & 1-\frac{2}{k+1}\frac{N}{F_Q} \\\hline
\widerow \tilde{Q} &\cbrm{N} & \bm{1} &\cbrm{F_Q} & -\frac{1}{N} & 1-\frac{2}{k+1}\frac{N}{F_Q} \\\hline
\widerow X & \brm{adj} & \brm{1} & \brm{1} & 0 & \frac{2}{k+1} \\\hline
\end{array}
\end{equation}
\caption{\em The matter content of the electric theory in the KSS model.\label{tab:KSSem}}
\end{table}
When there are also non-zero $t_i$ couplings, the $\U{1}_R$-symmetry is completely broken.  The $F$-term equation for the adjoint in this case can easily be solved by diagonalizing the $X$ using $\SU{N}$ rotations; the equation for a single entry $\chi$ on the diagonal is
\begin{equation}
W'=0\equiv\sum_{i=0}^{k-1}t_{i}\chi^{k-i}+\lambda.
\end{equation}
This is a $k$'th order polynomial so there are $k$ roots: hence
\begin{equation}
\tbrack{X}=\nbrack{\begin{array}{cccc}
\chi_1\mathbf{I}_{p_1}\\
& \chi_2\mathbf{I}_{p_2}\\
& & \ddots\\
&&& \chi_k\mathbf{I}_{p_k}\end{array}}\sep{where}\sum_{i=1}^kp_i=N.
\end{equation}
The gauge symmetry is broken as
\begin{equation}
\SU{N}\rightarrow\SU{p_1}\times\SU{p_2}\ldots\SU{p_k}\times\U{1}^{k-1}.
\end{equation}
Note that the parameter $\lambda$ is used as a Lagrange multiplier to fix $\Tr{}{X}=0$.

To get the corresponding magnetic theory we need to identify a set of elementary meson fields associated with composite operators of the electric model.  A crucial aspect of the superpotential is that it truncates the chiral ring; that is the equation of motion for $X$ (in for example the undeformed theory) sets $X^k=0$ (ignoring the Lagrange multiplier term) along the $F$-flat directions.  This means that when matching the moduli spaces, one need only consider operators up to $X^{k-1}$ (in the deformed theory $X^{k}$ is mapped to a polynomial in $X$ of order less than $k$).  Thus there are $k$ types of meson operator that we denote $m_j$;
\begin{equation}\label{eq:meson}
m_j=\tilde{Q}X^jQ,\hspace{5mm}j=0,\ldots,k-1.
\end{equation}
The $j=0$ object is the meson of usual Seiberg duality.  Indeed, the $k=1$ model is just the original Seiberg SQCD model if one integrates out the adjoint field.  
The field content of the magnetic theory is $q$, $\tilde{q}$, $m_j$ and $x$, where $x$ is an adjoint in the magnetic gauge group, and where the elementary magnetic mesons are directly and unambiguously identified with the composite operator $m_j$. 
Assuming for the moment that there is only the $t_0$ term in the electric theory then, as in Refs.\cite{KS:DKSS,KSS:DKSS}, the baryon matching implies that the gauge group of the full (unbroken) magnetic theory is
\begin{equation}
\SU{n}=\SU{kF_Q-N}\, .
\end{equation}
The matter content is summarised in Table \ref{tab:KSSmm}.
\begin{table}[!h]
\begin{equation}\nonumber
\begin{array}{|c|c|c|c|c|c|}\hline
\widerow & \SU{n} & \SU{F_Q}_L & \SU{F_Q}_R & \U{1}_B & \U{1}_R \\\hline
\widerow q & \brm{n} & \cbrm{F_Q} & \brm{1} & \frac{1}{n} & 1-\frac{2}{k+1}\frac{n}{F_Q} \\\hline
\widerow \tilde{q} &\cbrm{n} & \bm{1} &\brm{F_Q} & -\frac{1}{n} & 1-\frac{2}{k+1}\frac{n}{F_Q} \\\hline
\widerow x & \brm{adj} & \brm{1} & \brm{1} & 0 & \frac{2}{k+1} \\\hline
\widerow m_j & \brm{1} & \brm{F_Q} & \cbrm{F_Q} & 0 &  2-\frac{4}{k+1}\frac{N}{F_Q} + \frac{2j}{k+1} \\\hline
\end{array}
\end{equation}
\caption{\em The matter content of the magnetic theory in the KSS model; $n=kF_Q-N$.\label{tab:KSSmm}}
\end{table}
The superpotential in the deformed magnetic theory is of the form
\begin{equation}\label{eq:wmag}
W_{\mathrm{mag}}=-\sum_{i=0}^{k-1}\frac{t_i}{k+1-i}\Tr{}{x^{k+1-i}}+\frac{1}{\mu^2}\sum_{l=0}^{k-1}t_l\sum_{j=0}^{k-1-l}m_j\tilde{q}x^{k-1-j-l}q.
\end{equation}
This form of superpotential was deduced in Ref.\cite{KSS:DKSS} by equating the vacuum structures of the electric and magnetic theories.  The broken model in the magnetic theory is
\begin{equation}
\SU{n}\rightarrow\SU{F_Q-p_1}\times\SU{F_Q-p_2}\ldots\SU{F_Q-p_k}\times\U{1}^{k-1}.
\end{equation}
Note that the matching of the sub-theories $\SU{p_i}\leftrightarrow\SU{F_Q-p_i}$ is 
generically the standard SQCD duality.

Let us now turn to the anomaly matching of the deformed theories and, as we have proposed, introduce 
a singlet field $\phi$ whose role is to generate the $t_i$ couplings by acquiring a VEV: we therefore write $t_i\rightarrow\phi^{\rho_i}$ for some exponents $\rho_i$.  The coefficients in the magnetic superpotential are directly related to those in the electric one which, in turn, directly relates the singlet structure in the magnetic theory to that in the electric one. As a simple example consider the electric superpotential
\begin{equation}
W_{\mathrm{el}}=\phi^{\rho_0}\Tr{}{\frac{X^{k+1}}{k+1}}+\phi^{\rho_l}\Tr{}{\frac{X^{k+1-l}}{k+1-l}}.
\end{equation}
We will henceforth omit the Lagrange multiplier term, and also assume $k>2$ and $l>1$.  We may read off the magnetic superpotential from Eq.\eqref{eq:wmag}:
\begin{eqnarray}\label{eq:Wmag}
W_{\mathrm{mag}}&=&-\phi^{\rho_0}\Tr{}{\frac{x^{k+1}}{k+1}}-\phi^{\rho_l}\Tr{}{\frac{x^{k+1-l}}{k+1-l}}+\nonumber\\
&&\nbrack{\frac{\phi^{\rho_0}}{\mu^2}\sum_{j=0}^{k-1}m_j\tilde{q}x^{k-1-j}q+\frac{\phi^{\rho_l}}{\mu^2}\sum_{j=0}^{k-1-l}m_j\tilde{q}x^{k-1-j-l}q}.
\end{eqnarray}
Note that all the couplings of the magnetic theory, in particular the powers of $\phi$, are already determined by the vacuum structure.  However the parameter $\mu^2$ is an unknown free parameter that itself carries $R-$charge.  For example, it determines the matching conditions for the dynamical scales of the electric and magnetic theories ($\Lambda$ and $\bar{\Lambda}$ respectively) which are \cite{KS:DKSS}
\begin{equation}
\Lambda^{2N-F_Q}\bar{\Lambda}^{2n-F_Q}=\nbrack{\frac{\mu}{t_0}}^{2F_Q}.
\end{equation}
The $R$-symmetry is anomaly free (with respect to the mixed gauge-$R$ anomalies) so the dynamical scales do not carry $R$-charge or depend on $\phi$.  To be consistent the $R$-charge of $\mu$ has to be $\rho_0R_{\phi}$ and moreover $\mu$ must scale as $\mu\sim \phi^{\rho_0}$.  Both electric and magnetic theories now have an $R$-symmetry that is spontaneously broken by the VEV of $\phi$.

Indeed let us now lay out the charges of the fields explicitly. By virtue of the electric superpotential and the absence of mixed $\U{1}_R\times \SU{N}^2$ anomalies, the matter content of the electric theory is given by Table \ref{tab:DKSSem}.
\begin{table}[!h]
\begin{equation}\nonumber
\begin{array}{|c|c|c|c|c|c|}\hline
\widerow & \SU{N} & \SU{F_Q}_L & \SU{F_Q}_R & \U{1}_B & \U{1}_R \\\hline
\widerow Q & \brm{N} & \brm{F_Q} & \brm{1} & B_Q & 1-\frac{2N\nbrack{\rho_l-\rho_0}}{F_Q\sbrack{\rho_l\nbrack{1+k}-\rho_0\nbrack{1+k-l}}} \\\hline
\widerow \tilde{Q} &\cbrm{N} & \bm{1} &\cbrm{F_Q} & -B_Q & 1-\frac{2N\nbrack{\rho_l-\rho_0}}{F_Q\sbrack{\rho_l\nbrack{1+k}-\rho_0\nbrack{1+k-l}}} \\\hline
\widerow X & \brm{adj} & \brm{1} & \brm{1} & 0 & \frac{2\nbrack{\rho_l-\rho_0}}{\rho_l\nbrack{1+k}-\rho_0\nbrack{1+k-l}} \\\hline
\widerow \phi & \brm{1} & \brm{1} & \brm{1} & 0 & \frac{2l}{\rho_l\nbrack{1+k}-\rho_0\nbrack{1+k-l}} \\\hline
\end{array}
\end{equation}
\caption{\em The matter content of the electric theory in the deformed KSS model.  $B_{Q}$ is arbitrary (conventionally it is taken to be the baryon number, i.e.\ $B_Q=1/N$).\label{tab:DKSSem}}
\end{table}
Since we know the charges of the mesons from Eq.\eqref{eq:meson} we can independently derive the magnetic superpotential by writing down terms consistent with the $R$-symmetry.  This turns out to yield precisely $W_{\mathrm{mag}}$ as required, so that the superpotential deduced from the vacuum structure is consistent with that deduced from the $R$-symmetries.
At the classical level $W_{\mathrm{mag}}$ allows arbitrary values of $\rho_0$ and $\rho_l$.  However we will now see 
that the 't Hooft anomaly matching conditions constrain them.  

To do the anomaly matching we first need to identify correctly the elementary fields of the magnetic theory, and in particular deal with some arbitrariness about the definition of the elementary mesons. Specifically, the mesons'  role is to mirror (and project out via their equations of motion) the corresponding directions in the moduli space of the magnetic theory.   For example, consider the leading meson term of the magnetic superpotential.  Remembering that $\mu\sim\phi^{\rho_0}$ this can be written as
\begin{equation}
\phi^{-\rho_0}\sum_{j=0}^{k-1}m_j\tilde{q}x^{k-1-j}q\equiv\phi^{-\rho_0}\sum_{j=0}^{k-1}\nbrack{\tilde{Q}X^jQ}\nbrack{\tilde{q}x^{k-1-j}q}
\end{equation}
where we have exhibited the composite objects of the electric theory that are the erstwhile elementary mesons.  The equations of motion of $m_j$ therefore project out the corresponding directions in
the magnetic moduli space. The arbitrariness lies in the singlet content of what we call the elementary meson of the magnetic theory. Because the singlets carry $R$-charge, unless we determine this correctly, anomaly matching is impossible.

We therefore propose the following principle: the singlets are to be distributed among the elementary and composite magnetic mesons in the same way.  That is, rewriting the leading interaction as
\begin{equation}
\phi^{-\rho_0}\sum_{j=0}^{k-1}m_j\tilde{q}x^{k-1-j}q\equiv\sum_{j=0}^{k-1}\nbrack{\tilde{Q}(\phi^{\alpha}X)^jQ}\nbrack{\tilde{q}(\phi^{\alpha}x)^{k-1-j}q}\, ,
\end{equation}
where
\begin{equation}\label{eq:DKSSalpha}
\alpha=-\frac{\rho_0}{k-1},
\end{equation}
allows us to identify the elementary mesons $M_j$ of the magnetic theory as
\begin{equation}
M_j\equiv \tilde{Q}(\phi^{\alpha}X)^jQ.
\end{equation}
As we have said, identifying this singlet contribution to the meson for non-zero $\rho_0$ is an important step since it fixes the $R-$charges of the elementary mesons and hence the anomaly matching.

An independent check of this identification of the magnetic mesons is that the same identification is appropriate for the mapping between electric and magnetic baryons.  In detail, the baryons of the electric theory in the deformed KSS model are given by
\begin{equation}
B_{\{r_j\}}\sim\prod_{j=0}^{k-1}(X^jQ)^{r_j}=Q^NX^{\sum_{j=0}^{k-1}jr_j}\sep{where}\sum_{j=0}^{k-1}r_j=N
\end{equation}
(for concision we have omitted colour indices, which are contracted with a rank $N$ epsilon tensor).  The magnetic baryons to which these are mapped were found (by mass flow and by equating the global charges including $R$-charge) in Refs.\cite{KS:DKSS,KSS:DKSS} to be
\begin{equation}
b_{\{r_j\}}\sim\frac{t_0{}^{kF_Q/2}}{(\mu^{2})^{-n/2}}\prod_{j=0}^{k-1}\nbrack{x^jq}^{F_Q-r_j}\sim\nbrack{\phi^{\rho_0}}^{\nbrack{kF_Q-2n}/2}q^nx^{\sum_{j=0}^{k-1}j\nbrack{F_Q-r_j}}.
\end{equation}
Using $\sum_{j=0}^{k-1}j\nbrack{F_Q-r_j}=\sum_{j=0}^{k-1}jr_j-\nbrack{k-1}F_Q$, we find that the $R$-charges of the baryons are equal if
\begin{equation}
NR_Q-nR_q+\nbrack{k-1}\nbrack{N-\frac{kF_Q}{2}}R_X+\rho_0\nbrack{n-\frac{kF_Q}{2}}R_{\phi}=0
\end{equation}
which is satisfied for any $\rho_0$ (the specific $R$-symmetry of the undeformed models \cite{KS:DKSS,KSS:DKSS} is, of course, equivalent to taking $\rho_0=0$).  Using Eq.\eqref{eq:DKSSalpha} one finds that this is equivalent to
\begin{equation}
NR_Q-nR_q+\nbrack{\alpha R_{\phi}+R_X}\nbrack{k-1}\nbrack{N-\frac{kF_Q}{2}}=0.
\end{equation}
In other words, the prefactor
\begin{equation}
\frac{t_0{}^{kF_Q/2}}{(\mu^{2})^{-n/2}}\sim\nbrack{\phi^{\rho_0}}^{\nbrack{kF_Q-2n}/2}
\end{equation}
which is required to match the $R$-charges of the electric and magnetic baryons, is 
equivalent to associating with each $X$ precisely the same factor $\phi^{\alpha}$ as we did
did for the mesons
\begin{eqnarray}
B_{\{r_j\}}&\sim&\prod_{j=0}^{k-1}\sbrack{(\phi^{\alpha}X)^jQ}^{r_j}=Q^N\nbrack{\phi^{\alpha}X}^{\sum_{j=0}^{k-1}jr_j}\nonumber\\
b_{\{r_j\}}&\sim&\prod_{j=0}^{k-1}\sbrack{(\phi^{\alpha}x)^jq}^{F_Q-r_j}=q^n\nbrack{\phi^{\alpha}x}^{\sum_{j=0}^{k-1}j\nbrack{F_Q-r_j}}.
\end{eqnarray}
All that remains is to determine the $\U{1}_B$ charges of the magnetic quarks. They can be fixed by demanding that the $\U{1}_B$ charges of the electric and magnetic baryons match up:
\begin{equation}
B_q=-B_{\tilde{q}}=\frac{N}{n}B_{Q}.
\end{equation}

We may now proceed to anomaly matching. The mixed anomalies are found to be 
\begin{eqnarray}
\U{1}_B\times\SU{F_Q}_L^2&:&NB_Q\nonumber\\
\U{1}_B\times\SU{F_Q}_R^2&:&-NB_Q\nonumber\\
\U{1}_R\times\SU{F_Q}_L^2&:&-\frac{2N^2\nbrack{\rho_l-\rho_0}}{F_Q\sbrack{\rho_l\nbrack{1+k}-\rho_0\nbrack{1+k-l}}}\nonumber\\
\U{1}_R\times\SU{F_Q}_R^2&:&-\frac{2N^2\nbrack{\rho_l-\rho_0}}{F_Q\sbrack{\rho_l\nbrack{1+k}-\rho_0\nbrack{1+k-l}}}\nonumber\\
\U{1}_B\times\U{1}_R^2 &:& 0 \nonumber \\
\U{1}_R\times\U{1}_B^2&:&-\frac{4N^2B_Q^2\nbrack{\rho_l-\rho_0}}{\rho_l\nbrack{1+k}-\rho_0\nbrack{1+k-l}}\nonumber \\
\U{1}_B &:& 0 \nonumber \\
\U{1}_R &:& \frac{ (2 N^2 + k + 3)(\rho_0-\rho_l)-l (\rho_0-2)}
{\sbrack{\rho_l\nbrack{1+k}-\rho_0\nbrack{1+k-l}}}\, ,
\end{eqnarray}
in both theories.  The last two correspond to mixed $\U{1}$-gravity anomalies -- there is a 
contribution from the diagrams with gravitinos in the loop that is obviously universal and has been ommitted.  Three of the cubic anomalies also match
\begin{eqnarray}
\SU{F_Q}_L^3&:&N\nonumber\\
\SU{F_Q}_R^3&:&-N\nonumber\\
\U{1}_B^3&:&0.
\end{eqnarray}
Up to this point the matching occurs for arbitrary $\rho_0$ and $\rho_l$.  
However, the final $\U{1}_R^3$ anomaly only matches for the three combinations of $\rho_0$ and $\rho_l$ that are presented in Table \ref{tab:DKSSrhosol}.
\begin{table}[!h]
\begin{equation}\nonumber
\begin{array}{|c|c||c|}\hline
\widerow \rho_0 & \rho_l & \U{1}_R^3 \\\hline\hline
\widerow 0 & \mbox{any} & N^2-1-\frac{16N^4}{F_Q^2\nbrack{1+k}^3}+\nbrack{N^2-1}\nbrack{\frac{2}{1+k}-1}^3+\nbrack{\frac{2l}{\rho_l\nbrack{1+k}}-1}^3 \\\hline
\widerow \mbox{any} & \nbrack{\frac{k-l-1}{k-1}}\rho_0 & -F_Q^{-2}\rho_0^{-3}\sbrack{2N^4\rho_0^3+F_Q^2\nbrack{[k-1]^3+3\rho_0[k-1]^2+3\rho_0^2[k-1]-\rho_0^3[N^2-2]}} \\\hline
\widerow \mbox{any} & \rho_0 & -\rho_0^{-3}\nbrack{\rho_0-2}^3 \\\hline
\end{array}
\end{equation}
\caption{\em Choices of the exponents $\rho_0$ and $\rho_l$ that realize anomaly matching in the deformed KSS model with singlet.\label{tab:DKSSrhosol}}
\end{table}
There is also the special case $n=N$ which satisfies the anomaly matching for any $\rho_0$ and $\rho_l$. Note that, because of its complexity, the existence of simple solutions to the matching of the $\U{1}^3_R$ anomalies is certainly not a foregone conclusion. 

In the first case the $R$-symmetry is clearly that of the original KSS model with the singlet acting to preserve it under the addition of a deformation, but it is interesting that there are two other consistent choices.  The second case
\begin{equation}
\rho_l=\nbrack{\frac{k-l-1}{k-1}}\rho_0
\end{equation}
is equivalent to
\begin{equation}
R_X=1\sep{and}R_{\phi}=\frac{1-k}{\rho_0}.
\end{equation}
The last case has $\rho_l=\rho_0$, ergo $R_X=0$.  It is therefore less interesting, since the singlet has simply introduced an $R$-symmetry of its own under which none of the matter superfields are charged, and will be discarded.

The simplest choice of exponents in the $\rho_0\neq0$ model (where $\rho_l=\rho_0-l\rho_0/\nbrack{k-1}$) is the superpotential
\begin{equation}\label{eq:Wpneq0}
W_{\rho_0\neq0}=\phi^{k-1}\Tr{}{\frac{X^{k+1}}{k+1}}+\phi^{k-1-l}\Tr{}{\frac{X^{k+1-l}}{k+1-l}}
\end{equation}
with $R_X=1$ and $R_{\phi}=-1$.  Note that as far as the anomalies are concerned, we are at liberty to choose $l=k+1$ which generates a constant in the superpotential
\begin{equation}
W_{\rho_0\neq0}=\phi^{k-1}\Tr{}{\frac{X^{k+1}}{k+1}}+\phi^{-2}
\end{equation}
when $\phi$ gets a VEV.  Such a term could conceivably be generated dynamically. In the $\rho_0=0$ model the $R$-symmetry is precisely that of the KSS model.  In this case $\rho_l$ is unconstrained and the superpotential is
\begin{equation}\label{eq:Wpeq0}
W_{\rho_0=0}=\Tr{}{\frac{X^{k+1}}{k+1}}+\phi^{\rho_l}\Tr{}{\frac{X^{k+1-l}}{k+1-l}}
\end{equation}
with $R_{\phi}=\frac{2l}{\rho_l(k+1)}$.  Clearly both cases allow for generally deformed superpotentials with more than two terms; i.e.\ the general $R$-symmetric superpotentials (with these particular $R$-charges) can be written
\begin{eqnarray}
W_{\rho_0\neq0}&=&\phi^{k-1}\Tr{}{\frac{X^{k+1}}{k+1}f\nbrack{\phi X}}\nonumber\\
W_{\rho_0=0}&=&\Tr{}{\frac{X^{k+1}}{k+1}f\nbrack{\phi^{\rho_l/l}X^{-1}}}
\end{eqnarray}
where $f$ is an arbitrary polynomial of order $<k+1$.

There are various limits that one can consider.  In the $\rho_0=0$ theory for example, one can continuously take the $\phi\rightarrow0$ limit.  This simply corresponds to turning off the deformation of the KSS model.  The magnetic superpotential is
\begin{equation}
W_{\mathrm{mag}}=\Tr{}{\frac{x^{k+1}}{k+1}}+\phi^{\rho_l}\Tr{}{\frac{x^{k+1-l}}{k+1-l}}+\frac{1}{\mu^2}\sum_{j=0}^{k-1}M_j\nbrack{\tilde{q}x^{k-1-j}q}+\frac{\phi^{\rho_l}}{\mu^2}\sum_{j=0}^{k-1-l}M_j\nbrack{\tilde{q}x^{k-1-j-l}q}
\end{equation}
and goes continuously to the undeformed model as required.  In the oppositie $\phi\rightarrow\infty$ limit, we may define a rescaled $\hat{X}=\phi^{\rho_l/\nbrack{k+1-l}}X$ and similarly rescaled magnetic adoints, so that the electric superpotential becomes
\begin{equation}
W_{\rho_0= 0}=\phi^{-\rho_l\nbrack{k+1}/\nbrack{k+1-l}}\Tr{}{\frac{\hat{X}^{k+1}}{k+1}}+\Tr{}{\frac{\hat{X}^{k+1-l}}{k+1-l}}
\end{equation}
and the magnetic superpotential is
\begin{eqnarray}
W_{\mathrm{mag}}&=&\phi^{-\rho_l\nbrack{k+1}/\nbrack{k+1-l}}\Tr{}{\frac{\hat{x}^{k+1}}{k+1}}+\Tr{}{\frac{\hat{x}^{k+1-l}}{k+1-l}}+\nonumber\\
&&\frac{\phi^{-\rho_l\nbrack{k+1}/\nbrack{k+1-l}}}{\hat{\mu}^2}\sum_{j=0}^{k-1}\hat{M}_j\nbrack{\tilde{q}\hat{x}^{k-1-j}q}+\frac{1}{\hat{\mu}^2}\sum_{j=0}^{k-1-l}\hat{M}_j\nbrack{\tilde{q}\hat{x}^{k-1-j-l}q}
\end{eqnarray}
where
\begin{equation}
\hat{M}_j=\tilde{Q}\hat{X}^jQ\sep{and}\hat{\mu}=\phi^{\rho_l/\nbrack{k+1-l}}\mu.
\end{equation}
Thus in the large $\phi$ limit, keeping $\hat{\mu}$ constant, we have KSS duality with $k\rightarrow k-l$.  In terms of the original scales this corresponds to dialing down the Landau poles with respect to $\mu$.  Taking, for example, $\bar{\Lambda}\sim\Lambda$, the matching relation tells us that
\begin{equation}
\Lambda=\frac{\mu}{t_0}\sim\phi^{-\rho_l/\nbrack{k+1-l}}\hat{\mu}
\end{equation}
in this case. We shall later consider the RG flow in the theory.

\section{Models with more adjoints\label{sec:Xmod}}

\noindent Now that we have demonstrated the use of singlets in electric/magnetic duality, we can go on to apply this technique to models for which the dualities are not currently known.  All of the models presented below have electric theories based on SQCD, with gauge group $\SU{N}$ and chiral quark flavour symmetry $\SU{F_Q}$.  We will try always to denote fields and parameters in upper case for the electric theory and in lower case for the magnetic theory.  In each case the magnetic gauge group is $\SU{n}$ for some value of $n$ which must be determined.

\subsection{Two adjoints\label{sec:2X}}

\noindent The first model we consider extends SQCD with two $\SU{N}$ adjoints $X$ and $Y$, and an $\SU{N}$ singlet $\phi$.  Electric/magnetic duality with two adjoints was first discussed by Brodie and Strassler in Refs.\cite{B:2Adj,BS:Theatre}, however the form of their model is much more restricted than that which we are about to present. As our electric superpotential we take
\be\label{eq:2XWE}
W_{\mathrm{el}}=\phi^{\rho_X}\Tr{}{\frac{X^{k_X+1}}{k_X+1}}+\phi^{\rho_Y}\Tr{}{\frac{Y^{k_Y+1}}{k_Y+1}}+\Tr{}{XY}.
\ee
for some $\rho_X$ and $\rho_Y$, which allows a non-anomalous, global symmetry group
\be
\SU{F_Q}_L\times\SU{F_Q}_R\times\U{1}_B\times\U{1}_R
\ee
where $\U{1}_R$ is a unique $R$-symmetry.  To be exhaustive, we will consider both positive and 
negative values of $\rho_X$ and $\rho_Y$ (with negative values possibly being generated dynamically).

The $F$-terms for the adjoints give
\begin{eqnarray}\label{eq:2XF}
\phi^{\rho_X}X^{k_X}+Y&=&0\nonumber\\
\phi^{\rho_Y}Y^{k_Y}+X&=&0.
\end{eqnarray}
These equations truncate the chiral ring for all non-zero values of $\phi$.  To see this, one uses the first equation to set $Y=-\phi^{\rho_X}X^{k_X}$ then substitutes into the second equation to find
\be\label{eq:2Xtrunc}
\nbrack{-1}^{k_Y}\phi^{\rho_Xk_Y+\rho_Y}X^{k_Xk_Y}=-X.
\ee
Hence truncation occurs at $X^{k_Xk_Y}$ or, equivalently, $Y^{k_Xk_Y}$.  

Eq.\eqref{eq:2Xtrunc} resembles the usual truncation equation in KSS duality (see \S\ref{sec:DKSS}).  It is natural to suppose therefore that the duality proceeds in the same way, but with the effective value of $k$
\be
k^*=k_Xk_Y.
\ee
This implies the meson structure
\be
m_j\sim\tilde{Q}X^jQ,\hspace{5mm}j=0,\ldots,k^*-1
\ee
and consequently a dual gauge group $\SU{n}$ with
\be\label{eq:2Xn}
n=k^*F_Q-N.
\ee
Note that here it is convenient to use one of the adjoints $X$ to be `active'; thus  $m_{ak_X+b}\sim \tilde{Q}X^{ak_X+b}Q \sim \tilde{Q}Y^a X^bQ$, where $a< k_Y$ and $b<k_X$.

Let us now investigate the theory a little more closely.  The non-anomalous global symmetries and the matter content of the electric theory are displayed in Table \ref{tab:2Xem}.
\begin{table}[!h]
\be\nonumber
\begin{array}{|c|c|c|c|c|c|}\hline
\widerow & \SU{N} & \SU{F_Q}_L & \SU{F_Q}_R & \U{1}_B & \U{1}_R \\\hline
\widerow Q & \brm{N} & \brm{F_Q} & \brm{1} & \frac{1}{N} & 1-\frac{N}{F_Q} \\\hline
\widerow \tilde{Q} &\cbrm{N} & \bm{1} &\cbrm{F_Q} & -\frac{1}{N} & 1-\frac{N}{F_Q} \\\hline
\widerow X & \brm{adj} & \brm{1} & \brm{1} & 0 & \frac{2\nbrack{\rho_Xk_Y+\rho_Y}}{\rho_X\nbrack{1+k_Y}+\rho_Y\nbrack{1+k_X}} \\\hline
\widerow Y & \brm{adj} & \brm{1} & \brm{1} & 0 & \frac{2\nbrack{\rho_Yk_X+\rho_X}}{\rho_X\nbrack{1+k_Y}+\rho_Y\nbrack{1+k_X}} \\\hline
\widerow \phi & \brm{1} & \brm{1} & \brm{1} & 0 & \frac{2\nbrack{1-k^*}}{\rho_X\nbrack{1+k_Y}+\rho_Y\nbrack{1+k_X}}\\\hline
\end{array}
\ee
\caption{\em The matter content of the electric theory in the two adjoint model.\label{tab:2Xem}}
\end{table}
If one assumes the dual gauge group Eq.\eqref{eq:2Xn}, the equivalent information for the magnetic theory is as given in Table \ref{tab:2Xmm}, with the full properties of the elementary magnetic mesons yet to be determined.  The magnetic $\U{1}_B$ charges are again fixed by the baryon map which we will discuss shortly.  Following KSS duality, Table \ref{tab:2Xmm} also assumes that the magnetic superpotential takes the form
\be\label{eq:2XWmag1}
W_{\mathrm{mag}}=\phi^{\rho_X}\Tr{}{\frac{x^{k_X+1}}{k_X+1}}+\phi^{\rho_Y}\Tr{}{\frac{y^{k_Y+1}}{k_Y+1}}+\Tr{}{xy}+\mbox{meson terms}.
\ee 
The gauge singlet $\phi$ does not interact with the gauge group so is oblivious to the duality and must retain 
the same charges throughout.
\begin{table}[!h]
\be\nonumber
\begin{array}{|c|c|c|c|c|c|}\hline
\widerow & \SU{n} & \SU{F_Q}_L & \SU{F_Q}_R & \U{1}_B & \U{1}_R \\\hline
\widerow q & \brm{n} & \cbrm{F_Q} & \brm{1} & \frac{1}{n} & 1-\frac{n}{F_Q} \\\hline
\widerow \tilde{q} &\cbrm{n} & \bm{1} &\brm{F_Q} & -\frac{1}{n} & 1-\frac{n}{F_Q} \\\hline
\widerow x & \brm{adj} & \brm{1} & \brm{1} & 0 & \frac{2\nbrack{\rho_Xk_Y+\rho_Y}}{\rho_X\nbrack{1+k_Y}+\rho_Y\nbrack{1+k_X}} \\\hline
\widerow y & \brm{adj} & \brm{1} & \brm{1} & 0 & \frac{2\nbrack{\rho_Yk_X+\rho_X}}{\rho_X\nbrack{1+k_Y}+\rho_Y\nbrack{1+k_X}} \\\hline
\widerow \phi & \brm{1} & \brm{1} & \brm{1} & 0 & \frac{2\nbrack{1-k^*}}{\rho_X\nbrack{1+k_Y}+\rho_Y\nbrack{1+k_X}}\\\hline
\end{array}
\ee
\caption{\em The matter content of the magnetic theory in the two adjoint model, sans elementary mesons.  $n=k^*F_Q-N$.\label{tab:2Xmm}}
\end{table}

Now let us turn to the elementary mesons of the magnetic theory. In \S\ref{sec:DKSS} we found that powers of $\phi$ appeared in their definition, and the same occurs here.  To calculate the appropriate powers of $\phi$ we once again write $M_j=\tilde{Q}\nbrack{\phi^{\alpha}X}^jQ$
for $j=0,\ldots,k^*-1$ and some $\alpha$ to be determined, and then consider the meson terms in the magnetic superpotential.  These are
\be
W_{\mathrm{meson}}=\sum_{j=0}^{k^*-1}M_j\tilde{q}\nbrack{\phi^{\alpha}x}^{k^*-1-j}q
\ee 
where the power of $x$ is again shorthand for the corresponding power of 
$x^{1\ldots  k_X}$ and $y^{1\ldots k_Y}$.
Recall that no explicit $\mu^2 $ parameter is required now since all the $R$-charge is 
carried by the singlets. 
Note that the magnetic mesons still mirror the electric mesons, hence both use the same value of $\alpha$.  Demanding that all of the terms have $R$-charge $+2$ leads to a unique solution for $\alpha$
\be\label{eq:2Xalpha}
\alpha=\frac{\rho_X+\rho_Yk_X}{1-k^*}.
\ee
The final form for the electric mesons is therefore
\be\label{eq:2Xmeson}
M_j=\tilde{Q}\nbrack{\phi^{\nbrack{\rho_X+\rho_Yk_X}/\nbrack{1-k^*}}X}^jQ.
\ee
Equivalently, we could have used $Y$ instead of $X$ as the active adjoint to describe the mesons, in which case we would have found
\be
M_j=\tilde{Q}\nbrack{\phi^{\nbrack{\rho_Y+\rho_Xk_Y}/\nbrack{1-k^*}}Y}^jQ.
\ee
As usual, these mesons are to be added into the magnetic theory as elementary fields.  Their properties are summarised in Table \ref{tab:2Xmesons}.
\begin{table}[!h]
\be\nonumber
\begin{array}{|c|c|c|c|c|c|}\hline
\widerow & \SU{n} & \SU{F_Q}_L & \SU{F_Q}_R & \U{1}_B & \U{1}_R \\\hline
\widerow M_j & \brm{1} & \brm{F_Q} & \cbrm{F_Q} & 0 & 2\nbrack{1+j-\frac{N}{F_Q}} \\\hline
\end{array}
\ee
\caption{\em The elementary mesons of the magnetic theory in the two adjoint model.\label{tab:2Xmesons}}
\end{table}

We can now test the duality with 't~Hooft anomaly matching.  As before the $\U{1}_B^3$, $\U{1}_B \times \U{1}_R^2$ and $\U{1}_B$ anomalies are all zero; henceforth we will take them as read. The non-zero mixed anomalies are
\begin{eqnarray}
\U{1}_B\times\SU{F_Q}_L^2&:&1\nonumber\\
\U{1}_B\times\SU{F_Q}_R^2&:&-1\nonumber\\
\U{1}_R\times\SU{F_Q}_L^2&:&-N^2/F_Q\nonumber\\
\U{1}_R\times\SU{F_Q}_R^2&:&-N^2/F_Q\nonumber\\
\U{1}_R\times\U{1}_B^2&:&-2\nonumber \\
\U{1}_R&:&-2-N^2+\frac{2\nbrack{1-k^*}}{\rho_X\nbrack{1+k_Y}+\rho_Y\nbrack{1+k_X}}.
\end{eqnarray}
They are the same in both the electric and magnetic theories.  The cubic anomalies all match:  the non-zero ones are
\begin{eqnarray}
\SU{F_Q}_L^3&:&N\nonumber\\
\SU{F_Q}_R^3&:&-N\nonumber\\
\U{1}_R^3&:&-2+N^2-2N^4/F_Q^2+3R_{\phi}-3R_{\phi}^2+R_{\phi}^3
\end{eqnarray}
with $R_{\phi}$ as given in Table \ref{tab:2Xem}.  As the anomalies match for all values of $\rho_X$ and $\rho_Y$, any values giving well defined $R$-charges are allowed, and there is no constraint coming from the $\U{1}_R^3$ anomaly. This is because the $XY$ terms mean that $R_X=2-R_Y$ and consequently the $X$ and $Y$ fermions have equal and opposite $R$-charge and don't contribute here.
The quark $R$-charges are independent of $\rho_X$ and $\rho_Y$ for the same reason (i.e. their $R$-charges are determined by the required absence of $\SU{n}^2\times \U{1}_R$ anomalies and this 
gets no contribution from $X$ and $Y$ either). The only field whose $R$-charge depends on 
$\rho_X$ and $\rho_Y$ is therefore the singlet which contributes the same in both 
electric and magnetic theories. Note that demanding a well defined $R$-charge for the singlet eliminates the solution $\rho_X=\rho_Y=0$.

The point $\phi=0$ is singular. Assuming $\rho_X$ and $\rho_Y$ are both positive and that therefore $\alpha$ is negative, then the $F$-term equations Eq.\eqref{eq:2XF} have solution $X=Y=0$ at $\phi=0$, and the chiral ring truncates trivially. The behaviour of the $\phi=0$ theory and no adjoint mass is not well understood, but it is obvious that here  
ultimately the electric theory flows to that corresponding to normal SQCD with no adjoints, since we can integrate out the adjoint fields $X$ and $Y$ below their (Dirac) mass. The assignment of mesons other than $M_0$ also breaks down at this point.
Correspondingly the $\phi=0$ theory is ill-defined in the magnetic superpotential, which contains 
\be
W_{\mathrm{meson}}=  \sum_{j=0}^{k^*-1} 
M_j\tilde{q}\nbrack{x\phi^\alpha }^{k^*-1-j}q\, ,
\ee 
and which therefore diverges. This is hardly surprising because we are using $\phi$ to 
encode both the 
couplings in the superpotential {\em and} the dimensionful parameter that defines the elementary mesons in the magnetic description (i.e. the equivalent of the $\mu^2$ parameter of the KSS model).  

\subsubsection{Baryon matching\label{sec:2Xbaryon}}

\noindent The form of the mesons Eq.\eqref{eq:2Xmeson} strongly suggests that the adjoints should also come with these powers of the singlets in the baryon map between electric and magnetic theories.  Applying this observation we construct electric baryons
\be
B_{\{r_j\}}=\prod_{j=0}^{k^*-1}\sbrack{\nbrack{\phi^{\alpha}X}^jQ}^{r_j}
\ee
and magnetic baryons
\be
b_{\{r_j\}}=\prod_{j=0}^{k^*-1}\sbrack{\nbrack{\phi^{\alpha}x}^jq}^{F_Q-r_j}
\ee
with $\sum_{j=0}^{k^*-1}r_j=N$ and $\alpha$ as in Eq.\eqref{eq:2Xalpha}.  
Again we are for  convenience using $X$ to be the `active' adjoint; thus the dressed quarks correspond as  $ X^{ak_X+b}Q \sim Y^a X^bQ$, where $a< k_Y$ and $b<k_X$.
Our experience with KSS duality then leads us to propose the baryon map $B_{\{r_j\}}\leftrightarrow b_{\{r_j\}}$.  This map is totally consistent with all global symmetries; in particular the powers of the singlets are precisely those needed to match the $R$-charges of the baryons.

\subsubsection{Generalising the mass term\label{sec:2Xgenm}}

\noindent It may not always be desirable for the electric superpotential Eq.\eqref{eq:2XWE} to contain the mass term $XY$.  We will therefore outline how this term can be generalised.  
In order to do this we shall consider the matching of the two theories in the absence of this term, and 
then deduce that many alternative terms would be consistent with the duality. These terms would not necessarily truncate the chiral ring in which case a simple classical matching of the moduli spaces would not be possible. Nevertheless the fact that all the 
other tests of the duality are still satisfied is compelling.

In the absence of the $XY$ term, Eq.\eqref{eq:2XF} are modified and the chiral ring does not truncate in the same way.  Furthermore, the $R$-symmetry is relaxed to
\be
R_Q=R_{\tilde{Q}}=1-\frac{N}{F_Q}\nbrack{R_X+R_Y-1},\hspace{5mm}R_X=\frac{2-\rho_XR_{\phi}}{k_X+1},\hspace{5mm}R_Y=\frac{2-\rho_YR_{\phi}}{k_Y+1}
\ee
for arbitrary $R_{\phi}$ in the electric theory and
\be
R_q=R_{\tilde{q}}=1-\frac{n}{F_Q}\nbrack{R_x+R_y-1},\hspace{5mm}R_x=\frac{2-\rho_XR_{\phi}}{k_X+1},\hspace{5mm}R_y=\frac{2-\rho_YR_{\phi}}{k_Y+1}
\ee
in the magnetic theory.  These facts lead us to reconsider the meson structure of the theory.

The new modification of Eq.\eqref{eq:2XF} means that we are generally unable to define an `active' adjoint as we did previously, so are forced to consider the more general meson
\be
M_{ab}=\tilde{Q}(\phi^{\alpha_X}X)^a(\phi^{\alpha_Y}Y)^bQ
\ee
with $a=0,\ldots,k_X-1$ and $b=0,\ldots,k_Y-1$.  The parameters $\alpha_X$ and $\alpha_Y$ are undetermined functions of $R_{\phi}$.  The meson terms in the magnetic superpotential now look like
\be
W_{\mathrm{meson}}=\sum_{a=0}^{k_X-1}\sum_{b=0}^{k_Y-1}M_{ab}\tilde{q}\nbrack{\phi^{\alpha_X}x}^{k_X-1-a}\nbrack{\phi^{\alpha_Y}y}^{k_Y-1-b}q.
\ee
Demanding that these terms all have $R$-charge 2 allows one to write $\alpha_Y=\alpha_Y\nbrack{\alpha_X,R_{\phi}}$ as a function of $\alpha_X$ and $R_{\phi}$, but we are \emph{not} able to remove the $R_{\phi}$ dependence of the $\alpha$'s.  We thus conclude that the meson terms in the magnetic superpotential are inconsistent with more than one $R$-symmetry.  Hence if we remove the $XY$ term from the electric superpotential we are forced to replace it with a different third term that yields a single, non-anomalous $R$-symmetry in order to maintain the duality.  The precise value of $R_{\phi}$ is completely arbitrary as long as it is non-zero.  We thus have a very large amount of freedom in choosing the third term in the electric superpotential.

Having fixed the $R$-symmetry and the meson structure there is still one free parameter left in the theory; $\alpha_X$.  The only anomalies that might be sensitive to $\alpha_X$ are those involving the $R$-symmetry.  It turns out that the linear $\U{1}_R$ and the mixed $\U{1}_R\times\SU{F_Q}^2$ anomaly matching does not depend on $\alpha_X$.  The cubic $R$-symmetry anomaly matching, however, depends quadratically on $\alpha_X$.  If we choose one of the two values of $\alpha_X$ that set the cubic $R$-symmetry to zero all anomalies therefore match.

Using generalised dressed quarks we can construct electric baryons
\be
B_{\{r_{ab}\}}=\prod_{a=0}^{k_X-1}\prod_{b=0}^{k_Y-1}\sbrack{\nbrack{\phi^{\alpha_X}X}^a\nbrack{\phi^{\alpha_Y}Y}^bQ}^{r_{ab}}
\ee
and magnetic baryons
\be
b_{\{r_{ab}\}}=\prod_{a=0}^{k_X-1}\prod_{b=0}^{k_Y-1}\sbrack{\nbrack{\phi^{\alpha_X}x}^a\nbrack{\phi^{\alpha_y}Y}^bq}^{F_Q-r_{ab}}
\ee
with $\sum_{a=0}^{k_X-1}\sum_{b=0}^{k_Y-1}r_{ab}=N$.  The baryon map $B_{\{r_{ab}\}}\leftrightarrow b_{\{r_{ab}\}}$ is then consistent with all global symmetries for any choice of $\alpha_X$.  The only aspect of the theory that now needs testing is the meson sector.  We have not established how (or even if) the chiral ring truncates when the $XY$ term is replaced by some other interaction, which makes it difficult to see whether the mesonic degrees of freedom match up on either side of the proposed duality.  However, the fact that all other tests of duality, including the highly non-trivial 't Hooft anomaly matching and the baryon matching (itself, even less trivial than it was in the presence of the $XY$ term) are still passed suggests that the mesons will match up.  It may be that we have to resort to investigating the quantum chiral ring, rather than the classical chiral ring we have considered thus far, to see this.

\subsection{Three adjoints\label{sec:3X}}

The electric theory of this model is similar to the electric theory of the two adjoint model in \S\ref{sec:2X}, but with an extra $\SU{N}$ adjoint $Z$ and an extended superpotential
\be
W_{\mathrm{el}}=\phi^{\rho_X}\Tr{}{\frac{X^{k_X+1}}{k_X+1}}+\phi^{\rho_Y}\Tr{}{\frac{Y^{k_Y+1}}{k_Y+1}}+\phi^{\rho_Z}\Tr{}{\frac{Z^{k_Z+1}}{k_Z+1}}+\Tr{}{XY}+\Tr{}{YZ}.
\ee
For arbitrary values of $\rho_Y$, the electric theory has no non-anomalous $R$-symmetry.  One must choose
\be\label{eq:3XrhoY}
\rho_Y=\frac{\rho_X\nbrack{1-k_Yk_Z}-\rho_Z\nbrack{1-k_Xk_Y}}{k_Z-k_X}
\ee
in order for the theory to posses one.  This is markedly different from the two adjoint case in which all $\rho$'s were independent.  With $\rho_Y$ fixed as above, the non-anomalous, global symmetry group is once again
\be
\SU{F_Q}_L\times\SU{F_Q}_R\times\U{1}_B\times\U{1}_R.
\ee

The $F$-terms for the adjoints give
\begin{eqnarray}\label{eq:3XF}
\phi^{\rho_X}X^{k_X}+Y&=&0\nonumber\\
\phi^{\rho_Y}Y^{k_Y}+X+Z&=&0\nonumber\\
\phi^{\rho_Z}Z^{k_Z}+Y&=&0.
\end{eqnarray}
One can use the first equation to set $Y=-\phi^{\rho_X}X^{k_X}$, then the second equation to set $Z=-X-\nbrack{-1}^{k_Y}\phi^{\rho_Xk_Y+\rho_Y}X^{k_Xk_Y}$.  Upon substitution, the third equation reduces down to
\be
\phi^{\rho_Z}\sbrack{-X-\nbrack{-1}^{k_Y}\phi^{\rho_Xk_Y+\rho_Y}X^{k_Xk_Y}}^{k_Z}=\phi^{\rho_X}X^{k_X}
\ee
i.e.\
\be
X^{k_Xk_Yk_Z}\sim {\cal O}\nbrack{X^{k_Xk_Yk_Z-k_Xk_Y+1}}.
\ee
As long as $\phi\neq0$ this implies that the chiral ring is truncated at $X^{k_Xk_Yk_Z}$ or, equivalently, $Z^{k_Xk_Yk_Z}$.  
It is not generally possible to use $Y$ as the `active' adjoint as one is unable to express both $X$ and $Z$ purely in terms of $Y$.

Using $X$ as the active adjoint (the following argument would be equivalent if we used $Z$ instead) we now follow the procedure of \S\ref{sec:2X}.  Defining
\be
k^*=k_Xk_Yk_Z
\ee
we ought to be able to use the dual gauge group $\SU{n}$ with
\be
n=k^*F_Q-N.
\ee
The mesons are
\be
M_j=\tilde{Q}\nbrack{\phi^{\alpha}X}^jQ,\hspace{5mm}j=0,\ldots,k^*-1
\ee
but with a new value of $\alpha$
\be
\alpha=\frac{\rho_Z\nbrack{k^*-k_X}-\rho_X\nbrack{k^*-k_Z}}{\nbrack{1-k^*}\nbrack{k_Z-k_X}}
\ee
The magnetic superpotential is
\begin{eqnarray}
W_{\mathrm{mag}}&=&\phi^{\rho_X}\Tr{}{\frac{x^{k_X+1}}{k_X+1}}+\phi^{\rho_Y}\Tr{}{\frac{y^{k_Y+1}}{k_Y+1}}+\phi^{\rho_Z}\Tr{}{\frac{z^{k_Z+1}}{k_Z+1}}\nonumber \\
&& \hspace{3cm}+\Tr{}{xy}+\Tr{}{yz}
+\sum_{j=0}^{k^*-1}M_j\tilde{q}\nbrack{\phi^{\alpha}x}^{k^*-1-j}q
\end{eqnarray}
as expected.  We summarise the matter content for the electric and magnetic theories in Tables \ref{tab:3Xem} and \ref{tab:3Xmm} respectively.  All $\U{1}$ charge assignments can be seen to be consistent with both the meson structure and the magnetic superpotential as required.
\begin{table}[!h]
\be\nonumber
\begin{array}{|c|c|c|c|c|c|}\hline
\widerow & \SU{N} & \SU{F_Q}_L & \SU{F_Q}_R & \U{1}_B & \U{1}_R \\\hline
\widerow Q & \brm{N} & \brm{F_Q} & \brm{1} & \frac{1}{N} & 1-\frac{2N\nbrack{\rho_Z-\rho_X}}{F_Q\sbrack{\rho_Z\nbrack{1+k_X}-\rho_X\nbrack{1+k_Z}}} \\\hline
\widerow \tilde{Q} &\cbrm{N} & \bm{1} &\cbrm{F_Q} & -\frac{1}{N} & 1-\frac{2N\nbrack{\rho_Z-\rho_X}}{F_Q\sbrack{\rho_Z\nbrack{1+k_X}-\rho_X\nbrack{1+k_Z}}} \\\hline
\widerow X & \brm{adj} & \brm{1} & \brm{1} & 0 & \frac{2\nbrack{\rho_Z-\rho_X}}{\rho_Z\nbrack{1+k_X}-\rho_X\nbrack{1+k_Z}} \\\hline
\widerow Y & \brm{adj} & \brm{1} & \brm{1} & 0 & \frac{2\nbrack{\rho_Zk_X-\rho_Xk_Y}}{\rho_Z\nbrack{1+k_X}-\rho_X\nbrack{1+k_Z}} \\\hline
\widerow Z & \brm{adj} & \brm{1} & \brm{1} & 0 & \frac{2\nbrack{\rho_Z-\rho_X}}{\rho_Z\nbrack{1+k_X}-\rho_X\nbrack{1+k_Z}} \\\hline
\widerow \phi & \brm{1} & \brm{1} & \brm{1} & 0 & -\frac{2\nbrack{k_Z-k_Y}}{\rho_Z\nbrack{1+k_X}-\rho_X\nbrack{1+k_Z}} \\\hline
\end{array}
\ee
\caption{\em The matter content of the electric theory in the three adjoint model.\label{tab:3Xem}}
\end{table}
\begin{table}[!h]
\be\nonumber
\begin{array}{|c|c|c|c|c|c|}\hline
\widerow & \SU{n} & \SU{F_Q}_L & \SU{F_Q}_R & \U{1}_B & \U{1}_R \\\hline
\widerow q & \brm{n} & \cbrm{F_Q} & \brm{1} & \frac{1}{n} & 1-\frac{2n\nbrack{\rho_Z-\rho_X}}{F_Q\sbrack{\rho_Z\nbrack{1+k_X}-\rho_X\nbrack{1+k_Z}}} \\\hline
\widerow \tilde{q} &\cbrm{n} & \bm{1} &\brm{F_Q} & -\frac{1}{n} & 1-\frac{2n\nbrack{\rho_Z-\rho_X}}{F_Q\sbrack{\rho_Z\nbrack{1+k_X}-\rho_X\nbrack{1+k_Z}}} \\\hline
\widerow x & \brm{adj} & \brm{1} & \brm{1} & 0 & \frac{2\nbrack{\rho_Z-\rho_X}}{\rho_Z\nbrack{1+k_X}-\rho_X\nbrack{1+k_Z}} \\\hline
\widerow y & \brm{adj} & \brm{1} & \brm{1} & 0 & \frac{2\nbrack{\rho_Zk_X-\rho_Xk_Y}}{\rho_Z\nbrack{1+k_X}-\rho_X\nbrack{1+k_Z}} \\\hline
\widerow z & \brm{adj} & \brm{1} & \brm{1} & 0 & \frac{2\nbrack{\rho_Z-\rho_X}}{\rho_Z\nbrack{1+k_X}-\rho_X\nbrack{1+k_Z}} \\\hline
\widerow \phi & \brm{1} & \brm{1} & \brm{1} & 0 & -\frac{2\nbrack{k_Z-k_Y}}{\rho_Z\nbrack{1+k_X}-\rho_X\nbrack{1+k_Z}} \\\hline
\widerow M_j & \brm{1} & \brm{F_Q} & \cbrm{F_Q} & 0 & 2R_Q+j\nbrack{R_X+\alpha R_{\phi}} \\\hline
\end{array}
\ee
\caption{\em The matter content of the magnetic theory in the three adjoint model, $n=k^*F_Q-N$.\label{tab:3Xmm}}
\end{table}

The 't~Hooft anomaly matching reveals another significant difference between the two adjoint and three adjoint models.  The non-zero mixed anomalies are
\begin{eqnarray}
\U{1}_B\times\SU{F_Q}_L^2&:&1\nonumber\\
\U{1}_B\times\SU{F_Q}_R^2&:&-1\nonumber\\
\U{1}_R\times\SU{F_Q}_L^2&:&-2N^2\nbrack{\rho_Z-\rho_X}/F_Q\sbrack{\rho_Z\nbrack{1+k_X}-\rho_X\nbrack{1+k_Z}}\nonumber\\
\U{1}_R\times\SU{F_Q}_R^2&:&-2N^2\nbrack{\rho_Z-\rho_X}/F_Q\sbrack{\rho_Z\nbrack{1+k_X}-\rho_X\nbrack{1+k_Z}}\nonumber\\
\U{1}_R\times\U{1}_B^2&:&-4\nbrack{\rho_Z-\rho_X}/\sbrack{\rho_Z\nbrack{1+k_X}-\rho_X\nbrack{1+k_Z}}\nonumber\\
\U{1}_R&:&\frac{\nbrack{\rho_X-2}k_Z-\nbrack{\rho_Z-2}k_X+\nbrack{\rho_X-\rho_Z}\nbrack{3+2N^2}}{\sbrack{\rho_Z\nbrack{1+k_X}-\rho_X\nbrack{1+k_Z}}}.
\end{eqnarray}
They are still the same in both the electric and magnetic theories.  The cubic flavour anomalies are
\begin{eqnarray}
\SU{F_Q}_L^3&:&N\nonumber\\
\SU{F_Q}_R^3&:&-N
\end{eqnarray}
in both theories; the $\U{1}_B^3$ anomaly is zero.  However, the cubic $R$-symmetry anomalies do not immediately match as per the single adjoint model.  Their full forms are formidable, but one can write the difference as
\be\label{eq:3XR3}
f_{3X}\nbrack{\rho,k}\sbrack{\rho_Z-\rho_X}\sbrack{\rho_Z\nbrack{1-k_X}-\rho_X\nbrack{1-k_Z}}\sbrack{\rho_Z\nbrack{k_X-k^*}-\rho_X\nbrack{k_Z-k^*}}
\ee
where
\be
f_{3X}\nbrack{\rho,k}=\frac{4k^*F_Q\nbrack{1-2k^*}\nbrack{k^*F_Q-2N}}{\nbrack{1-k^*}\sbrack{\rho_Z\nbrack{1+k_X}-\rho_X\nbrack{1+k_Z}}^3}
\ee
When Eq.\eqref{eq:3XR3} is equal to zero all cubic anomalies match.  Solving for $\rho_Z$ yields three possibilities;
\be\label{eq:3XrhoZ}
\rho_Z=\rho_X,\hspace{5mm}\rho_Z=\frac{k_Z-1}{k_X-1}\rho_X\sep{or}\rho_Z=\frac{k^*-k_Z}{k^*-k_X}\rho_X
\ee
which, using Eq.\eqref{eq:3XrhoY}, result in
\be\label{eq:3XrhoY2}
\rho_Y=-\rho_Xk_Y,\hspace{5mm}\rho_Y=\frac{k_Y-1}{k_X-1}\rho_X\sep{or}\rho_Y=-\frac{k_Yk^*-1}{k^*-k_X}\rho_X
\ee
respectively.  Any choice of the $\rho$'s satisfying both Eq.\eqref{eq:3XrhoZ}, Eq.\eqref{eq:3XrhoY2} and giving well defined $R$-charges is then allowed.  Note that demanding a well defined $R$-charge for the singlet once again eliminates the solution $\rho_X=\rho_Y=\rho_Z=0$.

Of particular interest is the second of these solutions.  Choosing $\rho_X=k_X-1$ results in
\be
\rho_X=k_X-1,\hspace{5mm}\rho_Y=k_Y-1,\hspace{5mm}\rho_Z=k_Z-1
\ee
where we have used Eq.\eqref{eq:3XrhoY} to calculate $\rho_Y$.  Furthermore, one finds
\begin{eqnarray}
R_Q=R_{\tilde{Q}}&=&1-N/F_Q\nonumber\\
R_X=R_Y=R_Z&=&1\nonumber\\
R_{\phi}&=&-1
\end{eqnarray}
and the simple solution $\alpha=-1$.  This solution will be useful when we move on to consider theories with more adjoints.

As final check, we can consider the baryon matching.  This is simple and proceeds exactly as in \S\ref{sec:2Xbaryon}.  It remains independent of the values of the $\rho$'s and the $k$'s.

The point $\phi=0$ is again worth noting. The three solutions of Eq.\eqref{eq:3XrhoZ} allow for four possibilities.  $\rho_X$ and $\rho_Z$ must always have the same sign but Eq.\eqref{eq:3XrhoY2} tells is that $\rho_Y$ can have either.  If $\rho_X,\rho_Y,\rho_Z>0$ or $\rho_X,\rho_Y,\rho_Z<0$ the $F$-term equations Eq.\eqref{eq:3XF} do not truncate the chiral ring for $\phi=0$.  The same is true for $\rho_X,\rho_Z>0$, $\rho_Y<0$.  However, if $\rho_X,\rho_Z<0$, $\rho_Y>0$ the second of Eq.\eqref{eq:3XF} gives $Z=-X$ at $\phi=0$.  The first and third of Eq.\eqref{eq:3XF} then only have finite field value solutions at $\phi=0$ if $X^{k_X}=X^{k_Z}=0$ and $Y=0$.  The chiral ring is thus truncated at the lowest of $k_X$ and $k_Z$.  The 
electric theory is effectively reduced down to the deformed KSS model.

\subsection{$F_X$ adjoints\label{sec:FX}}

\noindent After \S\ref{sec:2X} and \S\ref{sec:3X} we expect the same principle will be extendable to an electric theory containing $F_X$ adjoints $X_i$ with $i=1,\ldots,F_X$.  Rather than go through the whole process with undetermined $\rho_i$'s, as we did in the previous sections, we will explicitly demonstrate the duality for a set of carefully chosen $\rho_i$'s in which the algebra is greatly simplified.

We choose
\be
W_{\mathrm{el}}=\sum_{i=1}^{F_X}\phi^{k_i-1}\Tr{}{\frac{X_i^{k_i+1}}{k_i+1}}+\sum_{i=1}^{F_X-1}\Tr{}{X_iX_{i+1}}
\ee
which gives the global symmetry group
\be
\SU{F_Q}_L\times\SU{F_Q}_R\times\U{1}_B\times\U{1}_R.
\ee
and $F$-terms
\begin{eqnarray}\label{eq:FXF}
\phi^{k_1-1}X_1^{k_1}+X_2&=&0\nonumber\\
\phi^{k_i-1}X_i^{k_i}+X_{i-1}+X_{i+1}&=&0,\hspace{5mm}i\neq1,F_X\nonumber\\
\phi^{k_{F_X}-1}X_{F_X}^{k_{F_X}}+X_{F_X-1}&=&0.
\end{eqnarray}
One can use the first of these equations to write $X_2$ in terms of $X_1$.  The second equation with $i=2$ then allows $X_3$ to be expressed in terms of $X_1$.  One continues, using the second equation for increasing values of $i$, until all of the $X_i$ have been expressed in terms of $X_1$ alone.  The third equation then reduces down to
\be
X_1^{k^*}\sim {\cal O}\nbrack{X_1^{k^*-\nbrack{k^*/k_{F_X}}+\nbrack{k^*/k_{F_X}k_{F_{X-1}}k_{F_{X-2}}}}}
\ee
where
\be
k^*=\prod_{i=1}^{F_X}k_i.
\ee
The chiral ring is thus truncated at $X_1^{k^*}$.  

Using $X_1$ as the active adjoint we now follow the usual procedure and use the dual gauge group $\SU{n}$ with
\be
n=k^*F_Q-N.
\ee
The mesons are
\be
M_j=\tilde{Q}\nbrack{\phi^{-1}X}^jQ,\hspace{5mm}j=0,\ldots,k^*-1
\ee
and the magnetic superpotential is
\be
W_{\mathrm{mag}}=\sum_{i=1}^{F_X}\phi^{k_i-1}\Tr{}{\frac{x_i^{k_i+1}}{k_i+1}}+\sum_{i=1}^{F_X-1}\Tr{}{x_ix_{i+1}}+\sum_{j=0}^{k^*-1}M_j\tilde{q}\nbrack{\phi^{-1}x_1}^{k^*-1-j}q.
\ee
We summarise the matter content for the electric and magnetic theories in Tables \ref{tab:FXem} and \ref{tab:FXmm} respectively.  All $\U{1}$ charge assignments can be seen to be consistent with both the meson structure and the magnetic superpotential as required.
\begin{table}[!h]
\be\nonumber
\begin{array}{|c|c|c|c|c|c|}\hline
\widerow & \SU{N} & \SU{F_Q}_L & \SU{F_Q}_R & \U{1}_B & \U{1}_R \\\hline
\widerow Q & \brm{N} & \brm{F_Q} & \brm{1} & \frac{1}{N} & 1-\frac{N}{F_Q} \\\hline
\widerow \tilde{Q} &\cbrm{N} & \bm{1} &\cbrm{F_Q} & -\frac{1}{N} & 1-\frac{N}{F_Q} \\\hline
\widerow X_i & \brm{adj} & \brm{1} & \brm{1} & 0 & 1 \\\hline
\widerow \phi & \brm{1} & \brm{1} & \brm{1} & 0 & -1 \\\hline
\end{array}
\ee
\caption{\em The matter content of the electric theory in the $F_X$ adjoint model.\label{tab:FXem}}
\end{table}
\begin{table}[!h]
\be\nonumber
\begin{array}{|c|c|c|c|c|c|}\hline
\widerow & \SU{n} & \SU{F_Q}_L & \SU{F_Q}_R & \U{1}_B & \U{1}_R \\\hline
\widerow q & \brm{n} & \cbrm{F_Q} & \brm{1} & \frac{1}{n} & 1-\frac{n}{F_Q} \\\hline
\widerow \tilde{q} &\cbrm{n} & \bm{1} &\brm{F_Q} & -\frac{1}{n} & 1-\frac{n}{F_Q} \\\hline
\widerow x_i & \brm{adj} & \brm{1} & \brm{1} & 0 & 1 \\\hline
\widerow \phi & \brm{1} & \brm{1} & \brm{1} & 0 & -1 \\\hline
\widerow M_j & \brm{1} & \brm{F_Q} & \cbrm{F_Q} & 0 & 2\nbrack{1+j-\frac{N}{F_Q}} \\\hline
\end{array}
\ee
\caption{\em The matter content of the magnetic theory in the $F_X$ adjoint model, $n=k^*F_Q-N$.\label{tab:FXmm}}
\end{table}

The 't Hooft anomaly matching is straightforward, thanks to the simple $R$-charges of the adjoints.  Indeed, the adjoints do not contribute to any of the anomalies we need to calculate as their fermionic components are not charged under any of the global symmetries.  We find non-zero mixed anomalies
\begin{eqnarray}
\U{1}_B\times\SU{F_Q}_L^2&:&1\nonumber\\
\U{1}_B\times\SU{F_Q}_R^2&:&-1\nonumber\\
\U{1}_R\times\SU{F_Q}_L^2&:&-N^2/F_Q\nonumber\\
\U{1}_R\times\SU{F_Q}_R^2&:&-N^2/F_Q\nonumber\\
\U{1}_R\times\U{1}_B^2&:&-2\nonumber\\
\U{1}_R&:&-3-N^2.
\end{eqnarray}
and non-zero cubic anomalies
\begin{eqnarray}
\SU{F_Q}_L^3&:&N\nonumber\\
\SU{F_Q}_R^3&:&-N\nonumber\\
\U{1}_R^3&:&-9+N^2-2N^4/F_Q^2.
\end{eqnarray}
All anomalies are the same in both theories.  Furthermore, the baryon matching goes exactly as in \S\ref{sec:2Xbaryon}.

We used particular values for the $\rho$'s throughout this section but we expect the duality will still exist if we change these values.  In general one is be able to choose values for two of the $\rho$'s freely and still obtain a non-anomalous $R$-symmetry consistent with the superpotential.  't~Hooft anomaly matching can impose extra constraints on the two free $\rho$'s.  From considerations of theories with more than three adjoints, it appears that, for even $F_X$, there are no extra constraints.  However, for odd $F_X$ only one of these two $\rho$'s remains free.  The second `free' $\rho$ is expressed in terms of the first, there being three possible solutions.

This is the one of two significant differences between theories with odd and even $F_X$.  The second being the truncation of the chiral ring for $\phi=0$.
When $\phi=0$ the chiral ring truncates trivially if $F_X$ is even but does not truncate at all (with these particular values of the $\rho$'s) if $F_X$ is odd.  To see this one must notice that the second equation of Eq.\eqref{eq:FXF} for $\phi=0$ connects oddly/evenly labelled $X_i$'s to oddly/evenly labelled $X_i$'s.  The first equation always gives $X_2=0$ at $\phi=0$.  One can then use the second equation to set $X_i=0$ for all evenly labelled $X_i$.  The third equation always gives $X_{F_X-1}=0$ at $\phi=0$.  If $F_X$ is even, $F_X-1$ is odd and one can use the second equation to set $X_i=0$ for all oddly labelled $X_i$ as well.  However, if $F_X$ is odd we have no way of truncating the oddly labelled $X_i$. We expect that other choices of the $\rho$'s will allow a truncation to the deformed KSS model as in the 3-adjoint example discussed earlier.

\section{RG flow\label{sec:RGflow}}

\noindent It is interesting to investigate the renormalisation group (RG) flow of some of the models we have discussed.  We will use the $a$-maximisation theorem developed in 
Refs.~\cite{Anselmi:1996dd, Anselmi:1997ys, IW:aMax, Barnes:2004jj} to do this. 
This theorem was used in the context of the KSS models in Refs.~\cite{Kutasov:2003iy,Kutasov:2003ux}.
In a nutshell, it tells us that the exact superconformal $R$-symmetry maximises the central charge
\be\label{eq:amax}
a\propto3\Tr{}{R^3}-\Tr{}{R}
\ee
where the trace is taken over all of the fermions in the theory.  The $R$-symmetry used in Eq.\eqref{eq:amax} is the most general, non-anomalous $R$-symmetry in the theory which commutes with charge conjugation.  Furthermore, the value of $a$ should decrease as the theory flows into the infrared.

Using this theorem, our general approach will be thus:
\begin{enumerate}
\item Start at $W=0$.
\item Use the $a$-maximisation theorem to find the exact superconformal $R$-symmetry.
\item\label{it:def} Write down all relevant deformations to the superpotential $\Delta W$ consistent with the global symmetries.  These are the gauge invariant operators with dimension $<3$, i.e.\ $R$-charge $<2$.
\item\label{it:fpts} Determine which of these deformations independently lead to consistent, non-trivial fixed points.  Consistency can only be achieved if $R_{\phi}\geq2/3$ at the proposed fixed point, with the inequality saturated if and only if $\phi$ is a free field and therefore absent from the superpotential.
\item Repeat steps \ref{it:def} and \ref{it:fpts} about all consistent fixed points.
\end{enumerate}
For step \ref{it:fpts} we are using the fact that, if $R_{\phi}<2/3$ at a fixed point, the singlet violates the unitarity bound; it is a gauge invariant spin-0 operator with dimension $<1$.  According to 
Ref.\cite{KS:DKSS} the solution to this problem is that $\phi$ is actually a free field charged under some accidental $\U{1}$ symmetry.  This accidental $\U{1}$ charge mixes with the existing singlet $R$-charge to set $R_{\phi}=2/3$.  However, if the deformation we found in step \ref{it:def} contained any powers of $\phi$ we arrive at a contradiction.  A free field cannot appear in the superpotential, and we conclude that such a fixed point cannot exist.

\subsection{Deformed KSS}

\noindent Consider the deformed KSS model of \S\ref{sec:DKSS}.  
In the absence of the superpotential, the general, non-anomalous $R$-symmetry is
\be
R_Q=R_{\tilde{Q}}=1-\hat{N}R_X\sep{where}\hat{N}=\frac{N}{F_Q}\, ,
\ee
and $R_X$ and $R_{\phi}$ are arbitrary.  Approximating the $\SU{N}$ gauge group by a $\U{N}$ gauge group and maximising $a$, one finds
\be\label{eq:amaxR}
R_X=\frac{10}{9+3\sqrt{20\hat{N}^2-1}},\hspace{5mm}R_{\phi}=\frac{2}{3}
\ee
provided $\hat{N}^2>1/2$.  When looking at the relevance of operators we will only be interested in the large $\hat{N}$ limit.  This is because any operator that is only relevant for sufficiently large $\hat{N}$ will actually be dangerously irrelevant for smaller $\hat{N}$: quark mass terms allow one to integrate out quarks, reduce $F_Q$ and therefore increase $\hat{N}$ until it is large enough for the operator in question to become relevant.  Note that working in the large $\hat{N}$ limit also legitimizes the approximation of  using  a $\U{N}$ gauge group, which greatly simplifies the algebra.

As $\hat{N}\rightarrow\infty$ in Eq.\eqref{eq:amaxR}, $R_X\rightarrow0$ and one finds several relevant deformations
\be\label{eq:RGdef}
\Delta W_0=\lambda_0X^{k_0+1},\hspace{5mm}\Delta W_1=\lambda_1\phi X^{k_1+1}\sep{and}\Delta W_2=\lambda_2\phi^2X^{k_2+1}.
\ee
There are other deformations not involving the adjoints, but these are of less interest so we will not discuss them.  We focus first on $W=\Delta W_2$, which fixes $R_{\phi}=1-\nbrack{k_2+1}R_X/2$.  $a$-maximisation results in $R$-charges of $R_X=0$ and $R_{\phi}=1$ for large $\hat{N}$.  $R_{\phi}>2/3$ so the fixed point is consistent and should exist.  Furthermore, one finds $\Delta W_0$ and $\Delta W_1$ are still relevant operators.  We thus try $W=\Delta W_2+\Delta W_1$.

With two terms in the superpotential the $R$-symmetry is completely fixed, with
\be
R_X=\frac{2}{2k_1-k_2+1}\sep{and}R_{\phi}=2-\frac{2\nbrack{k_1+1}}{2k_1-k_2+1}.
\ee
There are gauge singlets appearing in both terms so, going back to \S\ref{sec:DKSS}, this is the $\rho_0\neq0$, $\rho_l\neq0$ case.  If one wants a duality, Table \ref{tab:DKSSrhosol} tells us that we require either
\be
1=2\frac{k_1-1}{k_2-1}\sep{or}2=\frac{k_1-1}{k_2-1}
\ee
depending on whether $k_2>k_1$ (and $\rho_0=2$, $\rho_l=1$) or $k_2<k_1$ (and $\rho_0=1$, $\rho_l=2$).  Neither of these choices result in $R_{\phi}>2/3$ for non-zero values of both $k_2$ and $k_1$.  We conclude that a non-trivial fixed point for $W=\Delta W_2+\Delta W_1$ cannot exist in this class of electric/magnetic duals.

Instead, we try $W=\Delta W_2+\Delta W_0$.  Here one finds the fixed $R$-charges
\be
R_X=\frac{2}{k_0+1}\sep{and}R_{\phi}=\frac{k_0-k_2}{k_0+1}.
\ee
This superpotential falls into the $\rho_0=0$, $\rho_l\neq0$ category for which a duality is always possible as long as $k_0>k_2$.  One can easily show that, for $k_0>3k_2+2$, $R_{\phi}>2/3$.  A non-trivial fixed point for $W=\Delta W_2+\Delta W_0$ is therefore a consistent proposition.

\begin{figure}[!h]
\begin{center}
\includegraphics[width=50mm]{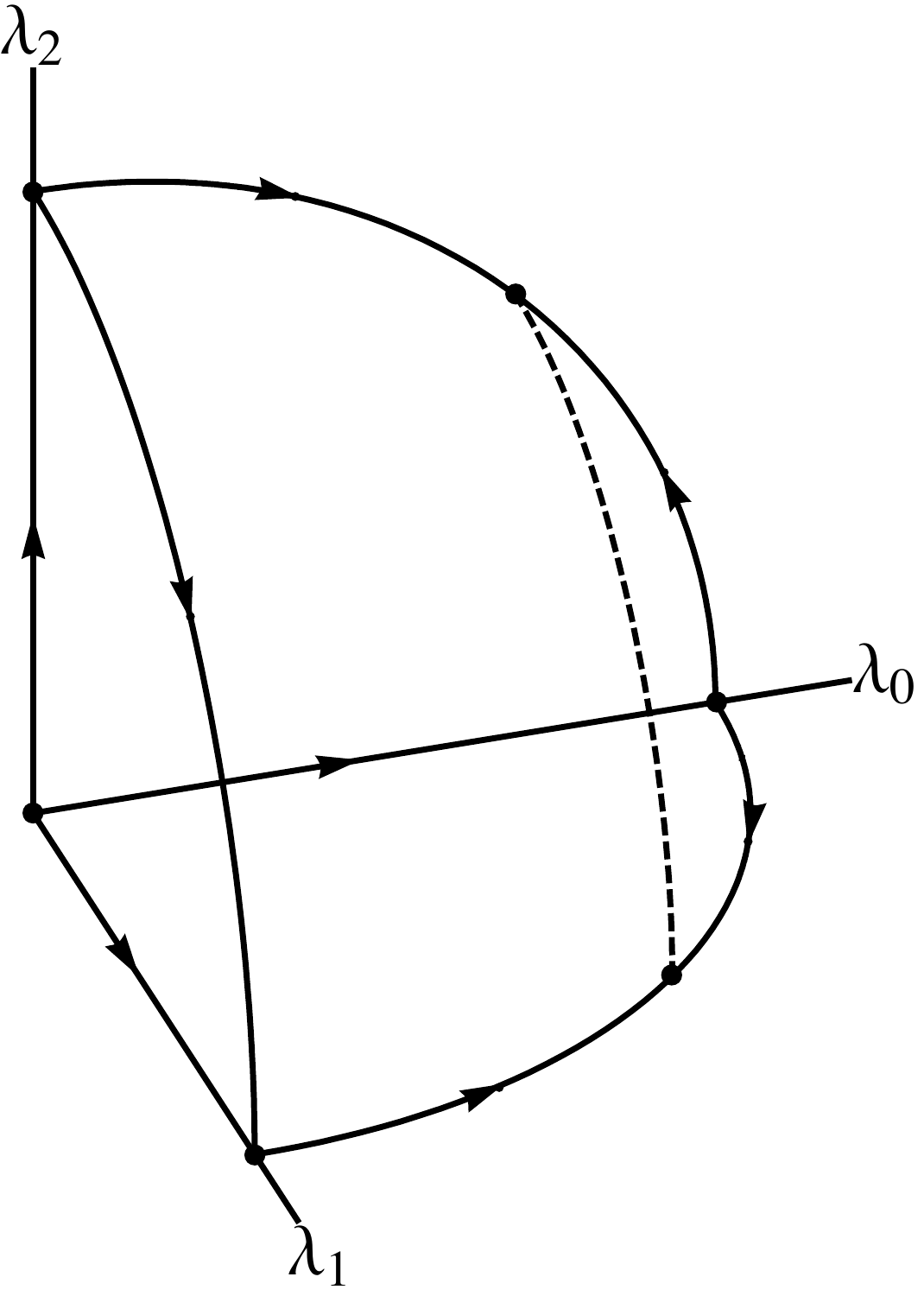}
\caption{\em The RG flow for the deformed KSS model of \S\ref{sec:DKSS} with the deformations Eq.\eqref{eq:RGdef} allowed by the duality.  $k_0>3k_2+2$ and $k_0>\frac{3}{2}k_1+\frac{1}{2}$ to ensure the $\lambda_2$-$\lambda_0$ and $\lambda_1$-$\lambda_0$ fixed points exist.  The flow along the dashed line connecting the two fixed points depends on the values of $k_2$ and $k_1$.\label{fig:RGsketch}}
\end{center}
\end{figure}

We have continued this analysis for all combinations of the deformations Eq.\eqref{eq:RGdef} maintaining the duality.  The results are summarised in Table \ref{tab:fixedpts} and the corresponding RG flow with the maximum number of fixed points is sketched in Figure \ref{fig:RGsketch}.
\begin{table}[!h]
\be\nonumber
\begin{array}{|c|c|c|c|c|c|c|}\hline
\widerow W & \multicolumn{2}{|c|}{+\Delta W_2} & \multicolumn{2}{|c|}{+\Delta W_1} & \multicolumn{2}{|c|}{+\Delta W_0} \\\hline
\widerow & \mbox{Relevant} & \mbox{Fixed Pt} & \mbox{Relevant} & \mbox{Fixed Pt} & \mbox{Relevant} & \mbox{Fixed Pt} \\\hline
\widerow 0 & \mbox{yes} & \mbox{yes} & \mbox{yes} & \mbox{yes} & \mbox{yes} & \mbox{yes} \\\hline
\widerow \Delta W_2 & - & - & \mbox{yes} & \mbox{no} & \mbox{yes} & k_0>3k_2+2 \\\hline
\widerow \Delta W_1 & \mbox{no} & \mbox{no} & - & - & \mbox{yes} & k_0>\frac{3}{2}k_1+\frac{1}{2}\\\hline
\widerow \Delta W_0 & k_2<\frac{1}{3}k_0+\frac{1}{3} & k_2<\frac{1}{3}k_0-\frac{2}{3} & k_1<\frac{2}{3}k_0+\frac{2}{3} & k_1<\frac{2}{3}k_0-\frac{1}{3} & - & - \\\hline
\end{array}
\ee
\caption{\em The fixed point analysis for the deformed KSS model of \S\ref{sec:DKSS} with the deformations Eq.\eqref{eq:RGdef} allowed by the duality.\label{tab:fixedpts}}
\end{table}
The flow along the dashed line in Figure \ref{fig:RGsketch} (connecting the $\lambda_2$-$\lambda_0$ and $\lambda_1$-$\lambda_0$ fixed points) depends on the values of $k_2$ and $k_1$.  The superpotential along this line is $W=\Delta W_2+\Delta W_1+\Delta W_0$, which only admits an $R$-symmetry if $k_2=\nbrack{k_1+k_0}/2$.  In this case the flow along the dashed line preserves $a$, suggesting a line of conformal fixed points.  If this is true we can use the reasoning of Ref.\cite{S:DualityCascade} to argue that the duality is \emph{exact} along the dashed line, rather than just an infrared duality.  When $k_2\neq\nbrack{k_1+k_0}/2$ there is no $R$-symmetry along the dashed line.  The difference in $a$ between the $\lambda_2$-$\lambda_0$ and $\lambda_1$-$\lambda_0$ fixed points is then
\begin{eqnarray}
\Delta a&=&a\nbrack{\lambda_2\mbox{-}\lambda_0}-a\nbrack{\lambda_1\mbox{-}\lambda_0}\\
&=&\frac{2k_1-\nbrack{k_2+k_0}}{\nbrack{1+k_0}^3}\nbrack{8-11k_0+2k_0^2+18k_1-12k_0k_1+12k_1^2+9k_2-3k_0k_2+6k_1k_2+3k_2^2}.\nonumber
\end{eqnarray}
For $\Delta a>0$ the flow is from $\lambda_2$-$\lambda_0$ to $\lambda_1$-$\lambda_0$.  If $\Delta a<0$ the flow is from $\lambda_1$-$\lambda_0$ to $\lambda_2$-$\lambda_0$

\subsection{More adjoints}

In models with two adjoints it was shown in Ref.\cite{IW:RGFlowAdj} that all operators of order less than four in the adjoints are relevant around the $W=0$ fixed point.  The addition of gauge singlets does not change this result.  The only such operator appearing in the models of \S\ref{sec:2X} is the mass term $XY$.  If one considers deforming the superpotential to $W=\lambda XY$ the $R$-symmetry must satisfy $R_Y=2-R_X$ so the corresponding fermions have equal and opposite $R$-charges.  The contribution of the adjoints to $a$ is therefore completely cancelled, leaving the value of $R_X$ undetermined.  Despite this freedom the RG flow of the two adjoint model is not very interesting.  Indeed, a quick glance at Table \ref{tab:2Xem} shows that $R_{\phi}$ is always negative for non-trivial choices of superpotential.  We thus conclude that the interesting terms in the two adjoint model are usually irrelevant and certainly don't generate non-trivial fixed points.

This is generally the case for models with an even number of adjoints.  After the mass terms have been added, there will be one adjoint $X_j$ for each adjoint $X_i$ such that $R_{X_j}=2-R_{X_i}$.  Consequently, none of the $R_{X_i}$ will appear in $a$ and we will end up with the situation where all interesting deformations to the superpotential are irrelevant.  For models with an odd number of adjoints this is not the case.  After adding the mass terms the $R$-charge cancellations will proceed as before but there will always be one adjoint whose $R$-charge cannot be paired up in this way.  We will thus be left with a single, undetermined $R_{X_i}$ appearing in $a$.  The RG flow then proceeds similarly to the model with a single adjoint, i.e.\ the deformed KSS model.  It should therefore come as no surprise that, when $\phi=0$, the models with an even number of adjoints flow to SQCD, while the models with an odd number of adjoints appear to flow to the deformed KSS model.

Note that we are generally interested in the regions of moduli space where $\phi\neq0$ and the $R$-symmetry is spontaneously broken.  All of the constraints we have derived in this section use $R$-symmetry arguments so do not apply at these points, leaving much more freedom to deform the superpotential.

\section{Models with antisymmetrics\label{sec:asym}}

\noindent Building on the success with adjoints in \S\ref{sec:Xmod}, we now turn to models containing antisymmetric representations of the gauge group.  We will base these on the models of Ref.\cite{ILS:NewDualities}.  These models are similar in spirit to KS duality but have a more complicated meson and baryon structure.  Once again, all electric theories are based on SQCD with gauge group $\SU{N}$.  The magnetic gauge group will be $\SU{n}$ for some $n$ to be determined.

\subsection{Two antisymmetrics\label{sec:2A}}

\noindent We will start with two antisymmetrics $A$ and $B$ along with their conjugates $\tilde{A}$ and $\tilde{B}$.  The electric superpotential is
\be
W_{\mathrm{el}}=\phi^{\rho_A}\Tr{}{\frac{(A\tilde{A})^{k_A+1}}{k_A+1}}+\phi^{\rho_B}\Tr{}{\frac{(B\tilde{B})^{k_B+1}}{k_B+1}}+\phi^{\sigma}\nbrack{\Tr{}{A\tilde{B}}+\Tr{}{\tilde{A}B}}.
\ee
for some $\rho_A$, $\rho_B$ and $\sigma$.  The inclusion of $\sigma\neq0$ turns out to be necessary for baryon matching; we will discuss it in more detail in \S\ref{sec:2Abaryon}.  For now, we merely note that
\be\label{eq:2Asigma}
\sigma=-\frac{{\rho_A\nbrack{1+k_B}+\rho_B\nbrack{1+k_A}}}{2k_Ak_B}.
\ee
The non-anomalous, global symmetry group for arbitrary $\rho_A$, $\rho_B$ and $\sigma$ is
\be
\SU{F_Q}_L\times\SU{F_Q}_R\times\U{1}_B\times\U{1}_A\times\U{1}_R
\ee
where $\U{1}_R$ is a unique $R$-symmetry.

The $F$-terms for the antisymmetrics give
\begin{eqnarray}\label{eq:2AF}
\phi^{\rho_A}\tilde{A}(A\tilde{A})^{k_A}+\phi^{\sigma}\tilde{B}&=&0\nonumber\\
\phi^{\rho_A}(A\tilde{A})^{k_A}A+\phi^{\sigma}B&=&0\nonumber\\
\phi^{\rho_B}\tilde{B}(B\tilde{B})^{k_B}+\phi^{\sigma}\tilde{A}&=&0\nonumber\\
\phi^{\rho_B}(B\tilde{B})^{k_B}B+\phi^{\sigma}A&=&0.
\end{eqnarray}
One uses the first two equations to write $B$ and $\tilde{B}$ in terms of $A$ and $\tilde{A}$ then, upon substitution into the last two equations, one finds
\begin{eqnarray}\label{eq:2Atrunc}
\phi^{\nbrack{\rho_A-\sigma}\nbrack{2k_B+1}-\sigma+\rho_B}\tilde{A}(A\tilde{A})^{2k_Ak_B+k_A+k_B}&=&\tilde{A}\nonumber\\
\phi^{\nbrack{\rho_A-\sigma}\nbrack{2k_B+1}-\sigma+\rho_B}(A\tilde{A})^{2k_Ak_B+k_A+k_B}A&=&A.
\end{eqnarray}
If $\phi\neq0$ the chiral ring truncates at $\tilde{A}(A\tilde{A})^{2k_Ak_B+k_A+k_B}$ and $(A\tilde{A})^{2k_Ak_B+k_A+k_B}A$ or, equivalently with $B$ and $\tilde{B}$.  We will discuss the $\phi=0$ case later.

Once again, Eq.\eqref{eq:2Atrunc} looks similar to the truncation equations of the base models Ref.\cite{ILS:NewDualities} for $\phi\neq0$.  We therefore assume the duality proceeds in the same way, but using the effective value of $k$
\be
k^*=\frac{1}{2}\sbrack{\nbrack{2k_A+1}\nbrack{2k_B+1}-1}.
\ee
This implies the meson structure
\begin{eqnarray}\label{eq:2Ameson}
M_j&=&\tilde{Q}(\phi^{2\alpha}A\tilde{A})^jQ,\hspace{5mm}j=0,\ldots,k^*\nonumber\\
P_j&=&Q[\phi^{\alpha}\tilde{A}(\phi^{2\alpha}A\tilde{A})^j]Q,\hspace{5mm}j=0,\ldots,k^*-1\nonumber\\
\tilde{P}_j&=&\tilde{Q}[(\phi^{2\alpha}A\tilde{A})^j\phi^{\alpha}A]\tilde{Q},\hspace{5mm}j=0,\ldots,k^*-1
\end{eqnarray}
where we have learnt from our experiences with adjoints and included a $\phi^{\alpha}$ with each antisymmetric from the start.  The dual gauge group is $\SU{n}$ with
\be\label{eq:2An}
n=\nbrack{2k^*+1}F_Q-4k^*-N.
\ee

With a magnetic superpotential
\begin{eqnarray}
W_{\mathrm{mag}}&=&\phi^{\rho_A}\Tr{}{\frac{(a\tilde{a})^{k_A+1}}{k_A+1}}+\phi^{\rho_B}\Tr{}{\frac{(b\tilde{b})^{k_B+1}}{k_B+1}}+\phi^{\sigma}\nbrack{\Tr{}{a\tilde{b}}+\Tr{}{\tilde{a}b}}+\nonumber\\
&&\sum_{j=0}^{k^*-1}P_jq[\phi^{\alpha}\tilde{a}\nbrack{\phi^{2\alpha}a\tilde{a}}^{k^*-1-j}]q+
\sum_{j=0}^{k^*-1}\tilde{P}_j\tilde{q}[\nbrack{\phi^{2\alpha}a\tilde{a}}^{k^*-1-j}\phi^{\alpha}a]\tilde{q}+\nonumber\\
&&\sum_{j=0}^{k^*}M_j\tilde{q}\nbrack{\phi^{2\alpha}a\tilde{a}}^{k^*-j}q
\end{eqnarray}
one finds the matter has the symmetry properties laid out in Table \ref{tab:2Aem} for the electric theory and Table \ref{tab:2Amm} for the magnetic theory.  The value of $\alpha$ is determined by demanding that the meson terms in the magnetic superpotential have $R$-charge $+2$.  This yields
\be\label{eq:2Aalpha}
\alpha=\frac{\rho_A\nbrack{1+k^*}+\rho_B\nbrack{1+2k_A}}{2k_Ak^*}.
\ee
\begin{table}[!h]
\be\nonumber
\begin{array}{|c|c|c|c|c|c|c|}\hline
\widerow & \SU{N} & \SU{F_Q}_L & \SU{F_Q}_R & \U{1}_B & \U{1}_A & \U{1}_R \\\hline
\widerow Q & \brm{N} & \brm{F_Q} & \brm{1} & \frac{1}{N} & 0 & 1-\frac{N+2k^*}{F_Q\nbrack{1+k^*}} \\\hline
\widerow \tilde{Q} &\cbrm{N} & \bm{1} &\cbrm{F_Q} & -\frac{1}{N} & 0 & 1-\frac{N+2k^*}{F_Q\nbrack{1+k^*}} \\\hline
\widerow A & \brm{asym} & \brm{1} & \brm{1} & \frac{2}{N} & 1 & \frac{\rho_A\nbrack{1+2k_B+k_Bk^*}+\rho_B\nbrack{1+k^*}}{\nbrack{1+k^*}\sbrack{\rho_A\nbrack{1+k_B}+\rho_B\nbrack{1+k_A}}} \\\hline
\widerow \tilde{A} & \cbrm{asym} & \brm{1} & \brm{1} & -\frac{2}{N} & -1 & \frac{\rho_A\nbrack{1+2k_B+k_Bk^*}+\rho_B\nbrack{1+k^*}}{\nbrack{1+k^*}\sbrack{\rho_A\nbrack{1+k_B}+\rho_B\nbrack{1+k_A}}} \\\hline
\widerow B & \brm{asym} & \brm{1} & \brm{1} & \frac{2}{N} & 1 & \frac{\rho_B\nbrack{1+2k_A+k_Ak^*}+\rho_A\nbrack{1+k^*}}{\nbrack{1+k^*}\sbrack{\rho_A\nbrack{1+k_B}+\rho_B\nbrack{1+k_A}}} \\\hline
\widerow \tilde{B} & \cbrm{asym} & \brm{1} & \brm{1} & -\frac{2}{N} & -1 & \frac{\rho_B\nbrack{1+2k_A+k_Ak^*}+\rho_A\nbrack{1+k^*}}{\nbrack{1+k^*}\sbrack{\rho_A\nbrack{1+k_B}+\rho_B\nbrack{1+k_A}}} \\\hline
\widerow \phi & \brm{1} & \brm{1} & \brm{1} & 0 & 0 & -\frac{2k_Ak_Bk^*}{\nbrack{1+k^*}\sbrack{\rho_A\nbrack{1+k_B}+\rho_B\nbrack{1+k_A}}}\\\hline
\end{array}
\ee
\caption{\em The matter content of the electric theory in the two antisymmetric model.\label{tab:2Aem}}
\end{table}
\begin{table}[!h]
\be\nonumber
\begin{array}{|c|c|c|c|c|c|c|}\hline
\widerow & \SU{n} & \SU{F_Q}_L & \SU{F_Q}_R & \U{1}_B & \U{1}_A & \U{1}_R \\\hline
\widerow q & \brm{n} & \cbrm{F_Q} & \brm{1} & \frac{1}{n} & \frac{\nbrack{F_Q-2}k^*}{n} & 1-\frac{n+2k^*}{F_Q\nbrack{1+k^*}} \\\hline
\widerow \tilde{q} &\cbrm{n} & \bm{1} &\brm{F_Q} & -\frac{1}{n} & -\frac{\nbrack{F_Q-2}k^*}{n} & 1-\frac{n+2k^*}{F_Q\nbrack{1+k^*}} \\\hline
\widerow a & \brm{asym} & \brm{1} & \brm{1} & \frac{2}{n} & \frac{N-F_Q}{n} & \frac{\rho_A\nbrack{1+2k_B+k_Bk^*}+\rho_B\nbrack{1+k^*}}{\nbrack{1+k^*}\sbrack{\rho_A\nbrack{1+k_B}+\rho_B\nbrack{1+k_A}}} \\\hline
\widerow \tilde{a} & \cbrm{asym} & \brm{1} & \brm{1} & -\frac{2}{n} & -\frac{N-F_Q}{n} & \frac{\rho_A\nbrack{1+2k_B+k_Bk^*}+\rho_B\nbrack{1+k^*}}{\nbrack{1+k^*}\sbrack{\rho_A\nbrack{1+k_B}+\rho_B\nbrack{1+k_A}}} \\\hline
\widerow b & \brm{asym} & \brm{1} & \brm{1} & \frac{2}{n} & \frac{N-F_Q}{n} & \frac{\rho_B\nbrack{1+2k_A+k_Ak^*}+\rho_A\nbrack{1+k^*}}{\nbrack{1+k^*}\sbrack{\rho_A\nbrack{1+k_B}+\rho_B\nbrack{1+k_A}}} \\\hline
\widerow \tilde{b} & \cbrm{asym} & \brm{1} & \brm{1} & -\frac{2}{n} & -\frac{N-F_Q}{n} & \frac{\rho_B\nbrack{1+2k_A+k_Ak^*}+\rho_A\nbrack{1+k^*}}{\nbrack{1+k^*}\sbrack{\rho_A\nbrack{1+k_B}+\rho_B\nbrack{1+k_A}}} \\\hline
\widerow \phi & \brm{1} & \brm{1} & \brm{1} & 0 & 0 & -\frac{2k_Ak_Bk^*}{\nbrack{1+k^*}\sbrack{\rho_A\nbrack{1+k_B}+\rho_B\nbrack{1+k_A}}}\\\hline
\widerow M_j & \brm{1} & \brm{F_Q} & \cbrm{F_Q} & 0 & 0 & \frac{F_Q\nbrack{2+2j+2k^*}-2N-4k^*}{F_Q\nbrack{1+k^*}}\\\hline
\widerow P_j & \brm{1} & \brm{asym} & \brm{1} & 0 & -1 & \frac{F_Q\nbrack{3+2j+2k^*}-2N-4k^*}{F_Q\nbrack{1+k^*}}\\\hline
\widerow \tilde{P}_j & \brm{1} & \brm{1} & \cbrm{asym} & 0 & 1 & \frac{F_Q\nbrack{3+2j+2k^*}-2N-4k^*}{F_Q\nbrack{1+k^*}}\\\hline
\end{array}
\ee
\caption{\em The matter content of the magnetic theory in the two antisymmetric model, $n=\nbrack{2k^*+1}F_Q-4k^*-N$.\label{tab:2Amm}}
\end{table}

Moving on to 't Hooft anomaly matching the picture is similar to \S\ref{sec:Xmod}.  The non-zero mixed anomalies are
\begin{eqnarray}\label{eq:2Amixed}
\U{1}_B\times\SU{F_Q}_L^2&:&1\nonumber\\
\U{1}_B\times\SU{F_Q}_R^2&:&-1\nonumber\\
\U{1}_R\times\SU{F_Q}_L^2&:&-N\nbrack{N+2k^*}/F_Q\nbrack{1+k^*}\nonumber\\
\U{1}_R\times\SU{F_Q}_R^2&:&-N\nbrack{N+2k^*}/F_Q\nbrack{1+k^*}\nonumber\\
\U{1}_R\times\U{1}_B^2&:&-2\nbrack{1+2k_A}\nbrack{1+2k_B}/\nbrack{1+k^*}\nonumber\\
\U{1}_R\times\U{1}_A^2&:&-k^*N\nbrack{N-1}/\nbrack{1+k^*}\nonumber\\
\U{1}_R&:&-2-\frac{3k^*N+N^2}{1+k^*}-\frac{2k_Ak_Bk^*}{\nbrack{1+k^*}\sbrack{\rho_A\nbrack{1+k_B}+\rho_B\nbrack{1+k_A}}}.
\end{eqnarray}
They are the same in both the electric and magnetic theories.  The cubic flavour anomalies are
\begin{eqnarray}\label{eq:2Acubic}
\SU{F_Q}_L^3&:&N\nonumber\\
\SU{F_Q}_R^3&:&-N
\end{eqnarray}
in both theories; the $\U{1}_B^3$ and $\U{1}_A^3$ anomalies are zero.  As in the three adjoint model, the cubic $R$-symmetry anomalies are both very complicated expressions which do not immediately match.  One can write the difference as
\be\label{eq:2AR3}
f_{2A}\nbrack{\rho,k}\sbrack{\rho_B\nbrack{1+2k_A}+\rho_A\nbrack{1+k^*}}\sbrack{\rho_B\nbrack{1+k^*}+\rho_A\nbrack{1+2k_B}}
\ee
where
\be
f_{2A}\nbrack{\rho,k}=-\frac{3k_Ak_Bk^*\sbrack{F_Q\nbrack{1+2k^*}-1-4k^*}\sbrack{F_Q\nbrack{1+2k^*}-2N-4k^*}}{\nbrack{1+k^*}^3\sbrack{\rho_A\nbrack{1+k_B}+\rho_B\nbrack{1+k_A}}^2}.
\ee
When Eq.\eqref{eq:2AR3} is equal to zero all cubic anomalies match.  Solving for $\rho_B$ yields two possibilities;
\be\label{eq:2ArhoBsol}
\rho_B=-\frac{k^*-k_B}{k_A\nbrack{1+k^*}}\rho_A\sep{or}\rho_B=-\frac{1+k^*}{1+2k_A}\rho_A.
\ee
Note that $\rho_A=\rho_B=0$ is not allowed as the $R$-charge of the singlet is not well defined.  Substituting into Eq.\eqref{eq:2Asigma} one finds
\be\label{eq:2Asigmasol}
\sigma=-\frac{k^*}{2k_A\nbrack{1+k^*}}\rho_A\sep{or}\sigma=\frac{k^*}{2k_B\nbrack{1+2k_A}}\rho_A
\ee
respectively.



For $\phi=0$ there are two truncation possibilities.  Choosing $\rho_A>0$ one finds strictly negative values of $\rho_B$ from Eq.\eqref{eq:2ArhoBsol}.  With this in mind, the $\sigma>0$ solution of Eq.\eqref{eq:2Asigmasol} reduces the $F$-term equations Eq.\eqref{eq:2AF} to
\be
\tilde{B}(B\tilde{B})^{k_B}=(B\tilde{B})^{k_B}B=0
\ee
at $\phi=0$ and the chiral ring does not truncate for $A$ and $\tilde{A}$.  On the other hand, the chiral ring truncates trivially for the $\sigma<0$ solution of Eq.\eqref{eq:2Asigmasol}.  The first two equations force $B=\tilde{B}=0$ which, in conjunction with the last two equations, forces $A=\tilde{A}=0$.  If $\rho_A<0$ one finds $\rho_B>0$ and the situation is identical, but with $A$'s and $B$'s exchanged.

\subsubsection{Baryon matching\label{sec:2Abaryon}}

\noindent The baryon matching scheme of Ref.\cite{ILS:NewDualities}, in conjunction with our meson structure Eq.\eqref{eq:2Ameson} and Eq.\eqref{eq:2Aalpha}, suggest electric baryons of the form
\be
B_r=(\phi^{\alpha}A)^rQ^{N-2r}
\ee
and magnetic baryons
\be
b_r=(\phi^{\alpha}a)^{\nbrack{F_Q-2}k^*-r}q^{F_Q-N+2r}
\ee
with the baryon map $B_r\leftrightarrow b_r$.  This map is totally consistent with all global symmetries.  The powers of the singlets are precisely those needed to match the $R$-charges of the baryons.  However, this is only the case for the value of $\sigma$ set down in Eq.\eqref{eq:2Asigma}.

If we had left $\sigma$ as an undetermined parameter we would have found
\be
\alpha=\frac{\rho_A\nbrack{4-F_Q+4k_B}+\rho_B\nbrack{4-F_Q+4k_A}-2\rho_Bk_AF_Q-2\sigma k_B\nbrack{F_Q+2k_AF_Q-4K_A}}{2F_Qk^*}
\ee
when we constrained the meson terms in the magnetic superpotential to have $R$-charge $+2$.  Using this value of $\alpha$ in the baryon map, the $R$-charges do not generally match up.  Even worse, the difference is dependent on the value of $r$ used in the baryons.  Hence we are not free to choose $\sigma$ arbitrarily.  If we wish to construct a baryon map which satisfies $R$-charge matching independently of $r$ we are forced to choose $\sigma$ as in Eq.\eqref{eq:2Asigma}.

\subsection{Three antisymmetrics\label{sec:3A}}

\noindent We now add an extra antisymmetric $C$ and its conjugate $\tilde{C}$ to the electric theory of \S\ref{sec:2A}, leading to a superpotential
\begin{eqnarray}
W_{\mathrm{el}}&=&\phi^{\rho_A}\Tr{}{\frac{(A\tilde{A})^{k_A+1}}{k_A+1}}+\phi^{\rho_B}\Tr{}{\frac{(B\tilde{B})^{k_B+1}}{k_B+1}}+\phi^{\rho_C}\Tr{}{\frac{(C\tilde{C})^{k_C+1}}{k_C+1}}+\nonumber\\
&&\phi^{\sigma}\nbrack{\Tr{}{A\tilde{B}}+\Tr{}{\tilde{A}B}+\Tr{}{B\tilde{C}}+\Tr{}{\tilde{B}C}}.
\end{eqnarray}
We define
\be\label{eq:3Ak*}
k^*=\frac{1}{2}\sbrack{\nbrack{2k_A+1}\nbrack{2k_B+1}\nbrack{2k_C+1}-1}
\ee
at this point to simplify the following expressions.  Note that, in line with our previous experiences, this is the value of $k^*$ we expect.  For arbitrary values of $\rho_B$, the electric theory has no non-anomalous $R$-symmetry.  One must choose
\be\label{eq:3ArhoB}
\rho_B=\frac{\rho_C\nbrack{k_A+k_B+2k_Ak_B}-\rho_A\nbrack{k_B+k_C+2k_Bk_C}}{k_C-k_A}+2\nbrack{1+k_B}\sigma
\ee
in order for the theory to possess one.  Furthermore, one requires
\be\label{eq:3Asigma}
\sigma=-\frac{\rho_C\nbrack{k^*-k_A}-\rho_A\nbrack{k^*-k_C}}{2\nbrack{1+k^*}\nbrack{k_C-k_A}}
\ee
for the baryon map to work out, as discussed in \S\ref{sec:2Abaryon}.  With $\rho_B$ and $\sigma$ fixed, the non-anomalous, global symmetry group is
\be
\SU{F_Q}_L\times\SU{F_Q}_R\times\U{1}_B\times\U{1}_A\times\U{1}_R
\ee
as it was for the two antisymmetric model.

The $F$-terms for the antisymmetrics give
\begin{eqnarray}\label{eq:3AF}
\phi^{\rho_A}\tilde{A}(A\tilde{A})^{k_A}+\phi^{\sigma}\tilde{B}&=&0\nonumber\\
\phi^{\rho_A}(A\tilde{A})^{k_A}A+\phi^{\sigma}B&=&0\nonumber\\
\phi^{\rho_B}\tilde{B}(B\tilde{B})^{k_B}+\phi^{\sigma}(\tilde{A}+\tilde{C})&=&0\nonumber\\
\phi^{\rho_B}(B\tilde{B})^{k_B}B+\phi^{\sigma}(A+C)&=&0\nonumber\\
\phi^{\rho_C}\tilde{C}(C\tilde{C})^{k_C}+\phi^{\sigma}\tilde{B}&=&0\nonumber\\
\phi^{\rho_C}(C\tilde{C})^{k_C}C+\phi^{\sigma}B&=&0.
\end{eqnarray}
One can use the first two equations to write $B$ and $\tilde{B}$ in terms of $A$ and $\tilde{A}$, then the middle two equations to write $C$ and $\tilde{C}$ in terms of $A$ and $\tilde{A}$.  Finally, the last two equations can be employed to find
\begin{eqnarray}
\tilde{A}(A\tilde{A})^{k^*}&\sim&O\nbrack{\tilde{A}(A\tilde{A})^{k^*-2k_Ak_B-k_A-k_B}}\nonumber\\
(A\tilde{A})^{k^*}A&\sim&O\nbrack{(A\tilde{A})^{k^*-2k_Ak_B-k_A-k_B}A}.
\end{eqnarray}
As long as $\phi\neq0$ the chiral ring is truncated at $\tilde{A}(A\tilde{A})^{k^*}$ and $(A\tilde{A})^{k^*}A$.  Equivalently, we could truncate in exactly the same way using $C$ and $\tilde{C}$.  The $\phi=0$ case is discussed later.

Assuming $\phi\neq0$ and using $A$ as the active antisymmetric (the argument would be equivalent if we used $C$ instead) we proceed as usual.  The dual gauge group is $\SU{n}$ with
\be\label{eq:3An}
n=\nbrack{2k^*+1}F_Q-4k^*-N.
\ee
$k^*$ is defined is Eq.\eqref{eq:3Ak*}.  The mesons are
\begin{eqnarray}\label{eq:3Ameson}
M_j&=&\tilde{Q}(\phi^{2\alpha}A\tilde{A})^jQ,\hspace{5mm}j=0,\ldots,k^*\nonumber\\
P_j&=&Q[\phi^{\alpha}\tilde{A}(\phi^{2\alpha}A\tilde{A})^j]Q,\hspace{5mm}j=0,\ldots,k^*-1\nonumber\\
\tilde{P}_j&=&\tilde{Q}[(\phi^{2\alpha}A\tilde{A})^j\phi^{\alpha}A]\tilde{Q},\hspace{5mm}j=0,\ldots,k^*-1
\end{eqnarray}
with
\be\label{eq:3Aalpha}
\alpha=\frac{\rho_C\nbrack{k^*-k_A}-\rho_A\nbrack{k^*-k_C}}{2\nbrack{1+k^*}\nbrack{k_C-k_A}}=-\sigma.
\ee
The magnetic superpotential is
\begin{eqnarray}
W_{\mathrm{el}}&=&\phi^{\rho_A}\Tr{}{\frac{(a\tilde{a})^{k_A+1}}{k_A+1}}+\phi^{\rho_B}\Tr{}{\frac{(b\tilde{b})^{k_B+1}}{k_B+1}}+\phi^{\rho_C}\Tr{}{\frac{(c\tilde{c})^{k_C+1}}{k_C+1}}+\nonumber\\
&&\phi^{\sigma}\nbrack{\Tr{}{a\tilde{b}}+\Tr{}{\tilde{a}b}+\Tr{}{b\tilde{c}}+\Tr{}{\tilde{b}c}}+\nonumber\\
&&\sum_{j=0}^{k^*-1}P_jq[\phi^{\alpha}\tilde{a}\nbrack{\phi^{2\alpha}a\tilde{a}}^{k^*-1-j}]q+
\sum_{j=0}^{k^*-1}\tilde{P}_j\tilde{q}[\nbrack{\phi^{2\alpha}a\tilde{a}}^{k^*-1-j}\phi^{\alpha}a]\tilde{q}+\nonumber\\
&&\sum_{j=0}^{k^*}M_j\tilde{q}\nbrack{\phi^{2\alpha}a\tilde{a}}^{k^*-j}q.
\end{eqnarray}
We summarise the matter content for the electric and magnetic theories in Tables \ref{tab:3Aem} and \ref{tab:3Amm} respectively.  All $\U{1}$ charge assignments are consistent with both the meson structure and the magnetic superpotential as required.
\begin{table}[!h]
\be\nonumber
\begin{array}{|c|c|c|c|c|c|c|}\hline
\widerow & \SU{N} & \SU{F_Q}_L & \SU{F_Q}_R & \U{1}_B & \U{1}_A & \U{1}_R \\\hline
\widerow Q & \brm{N} & \brm{F_Q} & \brm{1} & \frac{1}{N} & 0 & 1-\frac{N+2k^*}{F_Q\nbrack{1+k^*}} \\\hline
\widerow \tilde{Q} &\cbrm{N} & \bm{1} &\cbrm{F_Q} & -\frac{1}{N} & 0 & 1-\frac{N+2k^*}{F_Q\nbrack{1+k^*}} \\\hline
\widerow A & \brm{asym} & \brm{1} & \brm{1} & \frac{2}{N} & 1 & \frac{\rho_C-\rho_A}{\rho_C\nbrack{1+k_A}-\rho_A\nbrack{1+k_C}} \\\hline
\widerow \tilde{A} & \cbrm{asym} & \brm{1} & \brm{1} & -\frac{2}{N} & -1 & \frac{\rho_C-\rho_A}{\rho_C\nbrack{1+k_A}-\rho_A\nbrack{1+k_C}} \\\hline
\widerow B & \brm{asym} & \brm{1} & \brm{1} & \frac{2}{N} & 1 & \frac{1}{1+k^*}+\frac{2\nbrack{\rho_Ck_A-\rho_Ak_C}}{\rho_C\nbrack{1+k_A}-\rho_A\nbrack{1+k_C}} \\\hline
\widerow \tilde{B} & \cbrm{asym} & \brm{1} & \brm{1} & -\frac{2}{N} & -1 & \frac{1}{1+k^*}+\frac{2\nbrack{\rho_Ck_A-\rho_Ak_C}}{\rho_C\nbrack{1+k_A}-\rho_A\nbrack{1+k_C}} \\\hline
\widerow C & \brm{asym} & \brm{1} & \brm{1} & \frac{2}{N} & 1 & \frac{\rho_C-\rho_A}{\rho_C\nbrack{1+k_A}-\rho_A\nbrack{1+k_C}} \\\hline
\widerow \tilde{C} & \cbrm{asym} & \brm{1} & \brm{1} & -\frac{2}{N} & -1 & \frac{\rho_C-\rho_A}{\rho_C\nbrack{1+k_A}-\rho_A\nbrack{1+k_C}} \\\hline
\widerow \phi & \brm{1} & \brm{1} & \brm{1} & 0 & 0 & -\frac{2\nbrack{k_C-k_A}}{\rho_C\nbrack{1+k_A}-\rho_A\nbrack{1+k_C}} \\\hline
\end{array}
\ee
\caption{\em The matter content of the electric theory in the three antisymmetric model.\label{tab:3Aem}}
\end{table}
\begin{table}[!h]
\be\nonumber
\begin{array}{|c|c|c|c|c|c|c|}\hline
\widerow & \SU{n} & \SU{F_Q}_L & \SU{F_Q}_R & \U{1}_B & \U{1}_A & \U{1}_R \\\hline
\widerow q & \brm{n} & \cbrm{F_Q} & \brm{1} & \frac{1}{n} & \frac{\nbrack{F_Q-2}k^*}{n} & 1-\frac{n+2k^*}{F_Q\nbrack{1+k^*}} \\\hline
\widerow \tilde{q} &\cbrm{n} & \bm{1} &\brm{F_Q} & -\frac{1}{n} & -\frac{\nbrack{F_Q-2}k^*}{n} & 1-\frac{n+2k^*}{F_Q\nbrack{1+k^*}} \\\hline
\widerow a & \brm{asym} & \brm{1} & \brm{1} & \frac{2}{n} & \frac{N-F_Q}{n} & \frac{\rho_C-\rho_A}{\rho_C\nbrack{1+k_A}-\rho_A\nbrack{1+k_C}} \\\hline
\widerow \tilde{a} & \cbrm{asym} & \brm{1} & \brm{1} & -\frac{2}{n} & -\frac{N-F_Q}{n} & \frac{\rho_C-\rho_A}{\rho_C\nbrack{1+k_A}-\rho_A\nbrack{1+k_C}} \\\hline
\widerow b & \brm{asym} & \brm{1} & \brm{1} & \frac{2}{n} & \frac{N-F_Q}{n} & \frac{1}{1+k^*}+\frac{2\nbrack{\rho_Ck_A-\rho_Ak_C}}{\rho_C\nbrack{1+k_A}-\rho_A\nbrack{1+k_C}} \\\hline
\widerow \tilde{b} & \cbrm{asym} & \brm{1} & \brm{1} & -\frac{2}{n} & -\frac{N-F_Q}{n} & \frac{1}{1+k^*}+\frac{2\nbrack{\rho_Ck_A-\rho_Ak_C}}{\rho_C\nbrack{1+k_A}-\rho_A\nbrack{1+k_C}} \\\hline
\widerow c & \brm{asym} & \brm{1} & \brm{1} & \frac{2}{n} & \frac{N-F_Q}{n} & \frac{\rho_C-\rho_A}{\rho_C\nbrack{1+k_A}-\rho_A\nbrack{1+k_C}} \\\hline
\widerow \tilde{c} & \cbrm{asym} & \brm{1} & \brm{1} & -\frac{2}{n} & -\frac{N-F_Q}{n} & \frac{\rho_C-\rho_A}{\rho_C\nbrack{1+k_A}-\rho_A\nbrack{1+k_C}} \\\hline
\widerow \phi & \brm{1} & \brm{1} & \brm{1} & 0 & 0 & -\frac{2\nbrack{k_C-k_A}}{\rho_C\nbrack{1+k_A}-\rho_A\nbrack{1+k_C}} \\\hline
\widerow M_j & \brm{1} & \brm{F_Q} & \cbrm{F_Q} & 0 & 0 & \frac{F_Q\nbrack{2+2j+2k^*}-2N-4k^*}{F_Q\nbrack{1+k^*}}\\\hline
\widerow P_j & \brm{1} & \brm{asym} & \brm{1} & 0 & -1 & \frac{F_Q\nbrack{3+2j+2k^*}-2N-4k^*}{F_Q\nbrack{1+k^*}}\\\hline
\widerow \tilde{P}_j & \brm{1} & \brm{1} & \cbrm{asym} & 0 & 1 & \frac{F_Q\nbrack{3+2j+2k^*}-2N-4k^*}{F_Q\nbrack{1+k^*}}\\\hline
\end{array}
\ee
\caption{\em The matter content of the magnetic theory in the three antisymmetric model, $n=\nbrack{2k^*+1}F_Q-4k^*-N$.\label{tab:3Amm}}
\end{table}

We now consider the 't Hooft anomaly matching.  The mixed anomalies are exactly the same as those of the two antisymmetric model Eq.\eqref{eq:2Amixed}, but with the new version of $k^*$ given by Eq.\eqref{eq:3Ak*} and
\begin{eqnarray}
\U{1}_R&:&-2-\frac{3k^*N+N^2}{1+k^*}--\frac{2\nbrack{k_C-k_A}}{\rho_C\nbrack{1+k_A}-\rho_A\nbrack{1+k_C}}.
\end{eqnarray}
The cubic anomalies are also equivalent with the exception of the $\U{1}_R^3$ anomaly; the flavour anomalies are
\begin{eqnarray}
\SU{F_Q}_L^3&:&N\nonumber\\
\SU{F_Q}_R^3&:&-N
\end{eqnarray}
in both theories; the $\U{1}_B^3$ and $\U{1}_A^3$ anomalies are zero.  Again, the cubic $R$-symmetry anomalies do not immediately match.  One can write the difference as
\be\label{eq:3AR3}
f_{3A}\nbrack{\rho,k}\sbrack{\rho_Ck_A-\rho_Ak_C}\sbrack{\rho_C\nbrack{k^*-k_A}-\rho_A\nbrack{k^*-k_C}}^2
\ee
where
\be
f_{3A}\nbrack{\rho,k}=-\frac{6\sbrack{F_Q\nbrack{1+2k^*}-1-4k^*}\sbrack{F_Q\nbrack{1+2k^*}-2N-4k^*}}{\nbrack{1+k^*}^2\sbrack{\rho_C\nbrack{1+k_A}-\rho_A\nbrack{1+k_C}}^3}.
\ee
Where Eq.\eqref{eq:3AR3} is equal to zero all cubic anomalies match.  Solving for $\rho_C$ yields two possibilities;
\be\label{eq:3ArhoCsol}
\rho_C=\frac{k_C}{k_A}\rho_A\sep{or}\rho_C=\frac{k^*-k_C}{k^*-k_A}\rho_A
\ee
which, using Eq.\eqref{eq:3ArhoB}, result in
\be\label{eq:3ArhoBsol}
\rho_B=-\frac{k^*-k_B}{k_A\nbrack{1+k^*}}\rho_A\sep{or}\rho_B=-\frac{k_B+2k_Bk^*+k^*}{k^*-k_A}\rho_A
\ee
respectively.  Any choice of the $\rho$s satisfying both Eq.\eqref{eq:3ArhoCsol}, Eq.\eqref{eq:3ArhoBsol} and giving well defined $R$-charges is allowed.  Note that demanding a well defined $R$-charge for the singlet once again eliminates the solution $\rho_A=\rho_B=\rho_C=0$.  Substituting into Eq.\eqref{eq:3Asigma} and Eq.\eqref{eq:3Aalpha} one finds
\be\label{eq:3Asigmasol}
\sigma=-\alpha=-\frac{k^*}{2k_A\nbrack{1+k^*}}\rho_A\sep{or}\sigma=-\alpha=0
\ee
respectively.

As another check, we consider the baryon matching.  This goes exactly as in \S\ref{sec:2Abaryon} for all allowed choices of $\rho_A$.



When $\phi=0$, we get a deformed version of the models in Ref.\cite{ILS:NewDualities}. Explicitly,
Eq.\eqref{eq:3ArhoCsol} and Eq.\eqref{eq:3ArhoBsol} show that $\rho_A$ and $\rho_C$ always have the same sign whereas $\rho_B$ always has the opposite sign.  $\sigma$ either has the opposite sign to $\rho_A$ and $\rho_C$ or is zero.  We must therefore consider the $F$-term equations Eq.\eqref{eq:3AF} for several possibilities.  If $\rho_A,\rho_C>0$ then $\rho_B<0$ and $\sigma\leq0$.  Eq.\eqref{eq:3AF} tells us that $B=\tilde{B}=0$ at $\phi=0$ but says nothing about $A$ or $C$, hence the chiral ring is not truncated.  If $\rho_A,\rho_C<0$ then $\rho_B>0$ and $\sigma\geq0$.  The middle two equations reduce down to $A=-C$ and $\tilde{A}=-\tilde{C}$ in this case.  The remaining equations of Eq.\eqref{eq:3AF} then only have finite field value solutions at $\phi=0$ if $\tilde{A}(A\tilde{A})^{k_A}=(A\tilde{A})^{k_A}A=0$ and $\tilde{A}(A\tilde{A})^{k_C}=(A\tilde{A})^{k_C}A=0$.  If $\sigma=0$ this solution also sets $B=\tilde{B}=0$ and the chiral ring is truncated at the lowest of $k_A$ and $k_C$; we are, as claimed, left with a deformed version of the models in Ref.\cite{ILS:NewDualities}.  However, if $\sigma>0$ $B$ does not appear in the $F$-term equations at $\phi=0$ and the chiral ring remains untruncated.

\subsection{$F_A$ antisymmetrics?\label{sec:FA}}

\noindent The similarities between the first of the solutions for $\rho_B$ and $\sigma$ in the two and three antisymmetric models (see Eq.\eqref{eq:2ArhoBsol}, Eq.\eqref{eq:2Asigmasol} and Eq.\eqref{eq:3ArhoBsol}, Eq.\eqref{eq:3Asigmasol} respectively) suggests a pattern which may be extended to an arbitrary number of antisymmetrics.  This was certainly the case in \S\ref{sec:FX} for models with adjoints.  Thus far, we have been unable to realise a duality for a model with $F_A$ antisymmetrics.

We believe such a duality does exist.  Indeed, if the electric theory for such a model were to have superpotential
\be
W_{\mathrm{el}}=\sum_{i=1}^{F_A}\phi^{\rho_i}\Tr{}{\frac{(A_i\tilde{A}_i)^{k_i+1}}{k_i+1}}+\phi^{\sigma}\sum_{i=1}^{F_A-1}\nbrack{\Tr{}{A_i\tilde{A}_{i+1}}+\Tr{}{\tilde{A}_iA_{i+1}}}
\ee
the truncation would occur for $\phi\neq0$ is a similar way to the models of \S\ref{sec:2A} and \S\ref{sec:3A} but with
\be
k^*=\frac{1}{2}\nbrack{\prod_{i=1}^{F_A}\nbrack{2k_i+1}-1}.
\ee
Unfortunately the expressions for $R$-charges, $\sigma$ e.t.c.\ quickly become very unwieldy.  A simple generalisation of Eq.\eqref{eq:3ArhoBsol} and Eq.\eqref{eq:3Asigmasol} does not work.

\section{A note on symmetrics}

With a little effort the entirety of \S\ref{sec:asym} is immediately applicable to models with symmetric representations of the gauge group instead of antisymmetrics  (as was the case for the models of Ref.\cite{ILS:NewDualities}).  The only differences will be as follows.
\begin{itemize}
\item The dual gauge group is $\SU{n}$ with
\be
n=\nbrack{2k^*+1}F_Q+4k^*-N
\ee
and $k^*$ defined as before.
\item The mesons $P_j$ and $\tilde{P}_j$ are in symmetric representations of the appropriate quark flavour groups instead of antisymmetric representations.
\item The electric baryons go like
\be
B_r=(\phi^{\alpha}S)^r\nbrack{Q^2}^{N-r}
\ee
and magnetic baryons go like
\be
b_r=(\phi^{\alpha}s)^{2\nbrack{F_Q+2}k^*-r}\nbrack{q^2}^{F_Q-N+r}.
\ee
This accounts for the fact that symmetric tensors cannot be contracted with a single $\epsilon$.  Instead one needs to use two.  The baryon map is still $B_r\leftrightarrow b_r$.
\item The $R$-charges of the electric and magnetic quarks are
\be
R_Q=1-\frac{N-2k^*}{F_Q\nbrack{1+k^*}}\sep{and}R_q=1-\frac{n-2k^*}{F_Q\nbrack{1+k^*}}
\ee
respectively.  The $R$-charges of the mesons and the $R$-symmetry anomalies are updated accordingly.
\end{itemize}
The gauge singlet sector is completely oblivious to whether we use
symmetric or antisymmetric representations of the gauge group in the
theory.  In particular, the values of $\alpha$, $\sigma$ and the
constrained $\rho$'s will remain unchanged.  We have checked this
explicitly for the two symmetric model and it should apply to the rest
of \S\ref{sec:asym}.

\section{Conclusions}

We have investigated the role that singlets can play in establishing new 
electric/magnetic Kutasov type dualities (i.e. with superpotentials). Our models 
are based on ${\cal N}=1$ SQCD with $F_Q$ flavours of quark and antiquark with 
additional fields in the adjoint, antisymmetric or symmetric representations of 
$\SU{N}$. The 
singlets generate $R$-symmetry violating couplings once they acquire VEVs, and 
these couplings in turn allow for a simple truncation of the chiral ring. Consequently 
a much wider class of dualities can be investigated. On the other hand 't~Hooft 
anomaly matching can be performed on the full $R$-symmetric theory and provides 
the usual acid test of the duality. 

The central observation required to make the 't~Hooft anomaly matching work was 
how to identify the correct singlet content of the  elementary mesons in the magnetic theory. 
Once this has been achieved, it is possible to find electric/magnetic duals 
with any number of adjoints or antisymmetric or symmetrics.

\subsection*{Acknowledgements}
We acknowledge Valya Khoze and Joerg Jaeckel for helpful comments and observations. 

\end{document}